\documentclass[fleqn,usenatbib]{mnras}

\usepackage{newtxtext,newtxmath}
\usepackage[T1]{fontenc}

\DeclareRobustCommand{\VAN}[3]{#2}
\let\VANthebibliography\thebibliography
\def\thebibliography{\DeclareRobustCommand{\VAN}[3]{##3}\VANthebibliography}

\usepackage{graphicx}	% Including figure files
\usepackage{amsmath}	% Advanced maths commands
\usepackage[utf8]{inputenc}
\newcommand{\CHEOPS}{\mbox{\it CHEOPS}}

\newcommand{\software}[1]{\mbox{\sc {#1}}}
\newcommand{\var}[1]{\mbox{\sf {#1}}}

\title[SPIRou spectroscopy of EBLM~J0113+31]{Fundamental effective temperature measurements for eclipsing binary stars -- III. SPIRou near-infrared spectroscopy and CHEOPS photometry of the benchmark G0V star EBLM~J0113+31}

\author[P. F. L. Maxted et al.]{
P. F. L. Maxted,$^{1}$\thanks{E-mail: p.maxted@keele.ac.uk}
N. J. Miller,$^{1}$
S. Hoyer,$^{2}$
V. Adibekyan,$^{3}$
S. G. Sousa,$^{3}$
N. Billot,$^{4}$
A. Fortier,$^{5,6}$  \newauthor
A. E. Simon,$^{5}$
A. Collier Cameron,$^{7}$
M. I. Swayne,$^{1}$
P. Gutermann,$^{2,8}$
A. H. M. J. Triaud,$^{9}$ \newauthor
J. Southworth,$^{1}$
Y. Alibert,$^{5}$
R. Alonso,$^{10,11}$
G. Anglada,$^{12,13}$
T. Bárczy,$^{14}$
D. Barrado y Navascues,$^{15}$ \newauthor
S. C. C. Barros,$^{3,16}$
W. Baumjohann,$^{17}$
M. Beck,$^{4}$
T. Beck,$^{5}$
W. Benz,$^{5,6}$
X. Bonfils,$^{18}$
A. Brandeker,$^{19}$ \newauthor
C. Broeg,$^{5,6}$
M. Buder,$^{20}$
J. Cabrera,$^{20}$
S. Charnoz,$^{21}$
C. Corral van Damme,$^{22}$
Sz. Csizmadia,$^{20}$ \newauthor
M. B. Davies,$^{23}$
M. Deleuil,$^{2}$
L. Delrez,$^{24,25}$
O. Demangeon,$^{3,16}$
B.-O. Demory,$^{6}$
D. Ehrenreich,$^{4}$ \newauthor
A. Erikson,$^{20}$ 
L. Fossati,$^{17}$
M. Fridlund,$^{26,27}$
D. Gandolfi,$^{28}$
M. Gillon,$^{24}$
M. Güdel,$^{29}$
K. Heng,$^{6,30}$ \newauthor
J. E. Hernández Leon,$^{31}$
K. G. Isaak,$^{32}$
L. L. Kiss,$^{33,34}$
J. Laskar,$^{35}$
A. Lecavelier des Etangs,$^{36}$ \newauthor
M. Lendl,$^{4}$
C. Lovis,$^{4}$
D. Magrin,$^{37}$
M. Munari,$^{38}$
V. Nascimbeni,$^{37}$
G. Olofsson,$^{19}$
R. Ottensamer,$^{39}$ \newauthor
I. Pagano,$^{38}$
E. Pallé,$^{10}$
G. Peter,$^{40}$
G. Piotto,$^{37,41}$
D. Pollacco,$^{30}$
D. Queloz,$^{42,43}$
R. Ragazzoni,$^{37,41}$ \newauthor
N. Rando,$^{22}$
H. Rauer,$^{20,44,45}$
I. Ribas,$^{12,13}$
N. C. Santos,$^{3,16}$
G. Scandariato,$^{38}$
D. Ségransan,$^{4}$ \newauthor
A. M. S. Smith,$^{20}$
M. Steinberger,$^{17}$
M. Steller,$^{17}$
Gy. M. Szabó,$^{34,46}$
N. Thomas,$^{5}$
S. Udry,$^{4}$ \newauthor
V. Van Grootel,$^{25}$
N. Walton$^{47}$\\
(Affiliations are listed at the end of the paper)
}

\date{Accepted XXX. Received YYY; in original form ZZZ}

\pubyear{2021}

\begin{document}
\label{firstpage}
\pagerange{\pageref{firstpage}--\pageref{lastpage}}
\maketitle

\begin{abstract}
EBLM~J0113+31 is moderately bright (V=10.1), metal-poor ([Fe/H]\,$\approx-0.3$) G0V star with a much fainter M~dwarf companion on a wide, eccentric orbit (=14.3 d). We have used near-infrared spectroscopy obtained with the SPIRou spectrograph to measure the semi-amplitude of the M~dwarf's spectroscopic orbit, and high-precision photometry of the eclipse and transit from the CHEOPS and TESS space missions to measure the geometry of this binary system. From the combined analysis of these data together with previously published observations we obtain the following model-independent masses and radii: $M_1 = 1.029 \pm 0.025 M_{\odot}$, $M_2 = 0.197 \pm 0.003 M_{\odot}$, $R_1 = 1.417  \pm 0.014  R_{\odot}$, $R_2 = 0.215  \pm 0.002 R_{\odot}$. Using $R_1$ and the parallax from Gaia EDR3 we find that this star's angular diameter is $\theta = 0.0745 \pm 0.0007 $\,mas. The apparent bolometric flux of the G0V star corrected for both extinction and the contribution from the M~dwarf ($<0.2$\,per~cent) is ${\mathcal F}_{\oplus,0} = (2.62\pm 0.05)\times10^{-9}$\,erg\,cm$^{-2}$\,s$^{-1}$. Hence, this G0V star has an effective temperature $T_{\rm eff,1} = 6124{\rm\,K} \pm 40{\rm \,K\,(rnd.)} \pm 10 {\rm \,K\,(sys.)}$. EBLM~J0113+31 is an ideal benchmark star that can be used for ``end-to-end'' tests of the stellar parameters measured by large-scale spectroscopic surveys, or stellar parameters derived from asteroseismology with PLATO. The techniques developed here can be applied to many other eclipsing binaries in order to create a network of such benchmark stars. 
\end{abstract}


%%%%%%%%%%%%%%%%%%%%%%%%%%%%%%%%%%%%%%%%%%%%%%%%%%

%%%%%%%%%%%%%%%%% BODY OF PAPER %%%%%%%%%%%%%%%%%%

\section{Introduction}

Benchmark stars have properties that have been directly and accurately measured to good precision. They play a fundamental role in stellar astrophysics because we can only ascertain the accuracy and reliability of stellar models by comparing their predictions to the observed properties of real stars. Benchmark stars can also be used to establish empirical relations between stellar properties, e.g. colour -- effective temperature ($T_{\rm eff}$) relations \citep{2013ApJ...771...40B,2021ApJ...922..163V, 2015MNRAS.454.2863H}, or equations to estimate the mass or radius of a main-sequence star from $T_{\rm eff}$, $\log g$ and [Fe/H] \citep{2010A&ARv..18...67T}. Empirical relations are particularly useful in cases where stellar structure models are known to be unreliable, e.g. for low-mass stars,  where stellar models tend to under-predict the radius and over-predict $T_{\rm eff}$ \citep{2013ApJ...776...87S, 2019A&A...626A..32C, 2014MNRAS.437.2831Z, 2006ApJ...644..475B}.
 
Considerable effort has put into calibrating the $T_{\rm eff}$ scale for FGK-type dwarf stars. In recent years, this effort has been partly driven by the need for accurate $T_{\rm eff}$ estimates for planet host stars in order to estimate their masses and radii using stellar models \citep{2015MNRAS.447..846B, 2009ApJ...701..154B}. Much of the progress in characterising exoplanets in recent years has been due to the improved precision in measuring stellar masses and radii \citep{2019AREPS..47..141J}.

Benchmark FGK-type stars are also essential to calibrate the level of systematic and random uncertainties in massive spectroscopic surveys such as RAVE \citep{2020AJ....160...83S}, the Gaia-ESO survey \citep{2012Msngr.147...25G}, LAMOST \citep{2012RAA....12..735D}, GALAH \citep{2018MNRAS.478.4513B}, etc.  \citep{2014A&A...566A..98B, 2015A&A...582A..49H, 2018RNAAS...2..152J}. These surveys aim to reconstruct the formation history of the Galaxy by studying the pattern of elemental abundances in stars as a function of their mass, age and kinematics. \citet{2019ARA&A..57..571J} in their recent comprehensive review of ``industrial scale'' stellar abundance measurements suggest that it is today not possible to know the temperature of a star better than an accuracy of 50\,K. This uncertainty has a direct impact on reliability of trends observed in stellar abundance patterns between different stellar populations. Errors in $T_{\rm eff}$ are the dominant source of uncertainty when estimating the mass, radius, composition and age of a star \citep{2018A&A...620A.168V, 2019ARA&A..57..571J}. 
 
Validation and calibration of $T_{\rm eff}$ estimates for FGK-type dwarf stars currently rely on angular diameter measurements for a small sample of very bright stars such as Procyon, $\tau $~Cet, 18~Sco, $\alpha$~Cen\,A, etc. \citep{2022A&A...658A..47K, 2015A&A...582A..49H, 2013ApJ...771...40B}. Repeated measurements of the angular diameter for the same star often show differences much larger than the quoted uncertainties, with systematic errors of 5\,per~cent or more being quite common. For example, the 15 repeated measurements provided in Table~9 of \citet{2022A&A...658A..47K} for 7 G-type dwarf stars require an additional ``external error'' of about 0.04\,mas to be added to the quoted uncertainties to achieve $\chi^2_r=1$ for a fit of a constant offset to these difference. \citet{2022ApJ...927...31T} show that current uncertainties on measured interferometric angular diameters and bolometric fluxes set a systematic uncertainty floor of  $\approx 2$\,per~cent in $T_{\rm eff}$ for solar-type exoplanet host stars,  i.e. $\pm 120$\,K at $T_{\rm eff}=6000$\,K. 

Very low-mass stars (VLMSs, $\loa 0.2 M_{\odot}$) are attractive targets for exoplanet studies because of the possibility to detect and characterise the atmospheres of terrestrial planets in the habitable zones of these stars \citep{2021A&A...645A.100S}. There are very few well-characterised VLMSs because they are intrinsically very faint and small. For example, the recent empirical  colour\,--\,$T_{\rm eff}$, colour\,--\,luminosity and colour\,--\,radius relations published by \citet{2012ApJ...757..112B} are based on a sample that contains only one star with a spectral type later than M4V \citep[$M\approx 0.2 M_{\odot}$,][]{2019ApJ...871...63M}.

The EBLM project \citep{2013A&A...549A..18T} aims to improve our understanding of VLMSs by studying  eclipsing binaries with low-mass companions that have been found by the WASP survey \citep{Pollacco06}. These eclipsing binaries typically have a late-F- to mid-G-type primary star with an M-dwarf that contributes $\ll1$\,per~cent of the flux at optical wavelengths. The light curves of these EBLM systems look very similar to those of transiting hot Jupiters, which are the main targets for the WASP survey. As a result, dozens of these EBLM systems have been identified in the WASP survey. The analysis of the light curve combined with a spectroscopic orbit for the primary star and an estimate for its mass provides a direct estimate for the mass and radius of the M-dwarf companion \citep{2019A&A...625A.150V, 2019A&A...626A.119G}. With very high quality photometry it is possible to detect the eclipse of the M-dwarf and, hence, its surface brightness relative to the primary star. This surface brightness ratio can then be used to infer $T_{\rm eff}$ for the M-dwarf given an estimate of $T_{\rm eff}$ for the primary star and a surface brightness -- $T_{\rm eff}$ relation for the stars, either empirical \citep{2021A&A...649A.109G} or based on stellar model atmospheres. The first results from an on-going programme to measure mass, radius and $T_{\rm eff}$ for the M-dwarf in a sample of EBLM systems using ultrahigh-precision photometry obtained as part of the guaranteed time observations (GTO) with the CHEOPS mission \citep{2021ExA....51..109B} have been published by \citet{2021MNRAS.506..306S}. Most of the targets for this programme were selected from a sample of over 100 EBLM systems with spectroscopic orbits published by \citet{2017A&A...608A.129T}. 

The first study of EBLM~J0113+31, the target for this study, was published by \citet[][GMC+2014 hereafter]{2014A+A...572A..50G}. They used ground-based photometry of the eclipse in the J-band to infer $T_{\rm eff} \approx 3900$\,K for the very low-mass companion, much higher than expected given their estimate for its mass ($M_2=0.186 \pm 0.010 M_{\odot}$). Subsequent analysis of the TESS light curve for this binary system by \citet{2020MNRAS.498L..15S} found no evidence for a very hot companion. Their value of $T_{\rm eff,2} = 3208 \pm 43$\,K is similar to that for other VLMSs. They conclude that the anomalous result from GMC+2014 was due to systematic errors in the J-band photometry, illustrating the need for very high-quality space-based photometry to make reliable measurements of $T_{\rm eff,2}$ in EBLM systems.

Here we present new photometry of the transit and eclipse in  EBLM~J0113+31 obtained with CHEOPS, and high-resolution, phase-resolved spectroscopy obtained with the near-infrared échelle spectrograph SPIRou on the Canada-France-Hawaii telescope. We have used the SPIRou spectroscopy to directly measure the semi-amplitude of the M-dwarf's spectroscopic orbit. We have used this measurement combined with the analysis of the new light curves and other published data to directly and accurately measure the mass, radius and $T_{\rm eff}$ of both stars in this binary system. We discuss the use of the techniques developed here to determine fundamental stellar properties for stars in EBLM systems, and conclude that it is now feasible to establish a network of well-studied EBLM systems that  will be an ideal set of benchmark stars for future surveys.

\section{Observations}

\subsection{CHEOPS photometry}
 CHEOPS is a telescope with an effective aperture of 30\,cm in low Earth orbit that is designed to obtain ultrahigh precision broadband photometry of bright stars \citep{2021ExA....51..109B}. To achieve this, the instrument has been purposely defocused to produce a point-spread function (PSF) with a diameter of approximately 32\arcsec. We observed two transits and one eclipse of EBLM~J0113+31 with CHEOPS (Table~\ref{tab:obslog}). There are gaps in the observations due to occultation of the target by the Earth and passages of the satellite though the South Atlantic Anomaly. 
 
The raw data were processed using version 13.1.0 of the CHEOPS data reduction pipeline \citep[DRP,][]{2020A&A...635A..24H}. The DRP corrects the images for environmental and instrumental effects before performing aperture photometry of the target. The contamination of the photometric aperture during the exposure by nearby stars is estimated using simulations of the field of view based on the  Gaia DR2 catalogue \citep{2018A&A...616A...1G}. The instrument response function for CHEOPS is very similar to the Gaia G band. The detector used on the CHEOPS instrument is a frame-transfer charge-coupled device (CCD), so the DRP must also account for the ``smear'' trails from bright stars produced during the frame transfer. Both of these effects (contamination and smear) vary from image to image because the satellite rotates continuously during its 99-minute orbit. 
 
Aperture photometry is extracted by the DRP using three different fixed aperture sizes labelled ``RINF'', ``DEFAULT'' and ``RSUP'' (at radii of 22.5, 25.0 and 30.0 pixels, respectively) and a further ``OPTIMAL'' aperture whose size is determined for each visit dependent upon the amount of contamination. The observed and processed data are made available on the Data Analysis Center for Exoplanets (DACE) web platform\footnote{The DACE platform is available at \url{http://dace.unige.ch}}. We downloaded our data from DACE using {\fontfamily{qcr}\selectfont pycheops}\footnote{\url{https://pypi.org/project/pycheops/}}, a \textsc{python} module developed for the analysis of data from the CHEOPS mission \citep{pycheops}.
 
\begin{figure}
	\includegraphics[width=\columnwidth]{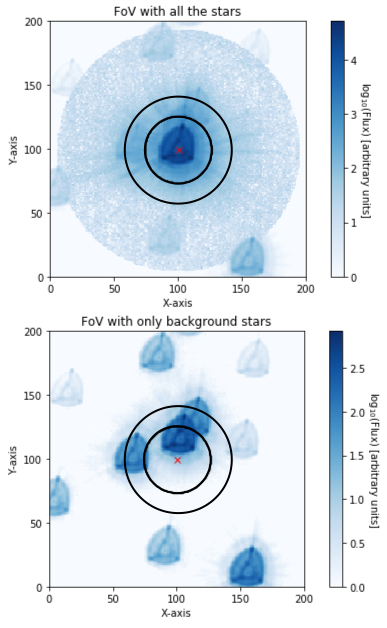}
    \caption{Simulated images of the CHEOPS field of view. Upper panel: all the stars in the field of view including the target. Lower panel: The target has been removed to show only the background stars in the field of view. Black circles show the DEFAULT (25 pixel) and OPTIMAL (40 pixel) apertures and the red cross shows the location of the target star.}
    \label{fig:fov}
\end{figure}

There are three stars that are 5--6 magnitudes fainter than EBLM~J0113+31 within 1\arcmin\ of the target (Fig.~\ref{fig:fov}). As a result, the OPTIMAL aperture is set to its maximum allowed value by the DRP (40 pixels = 40\arcsec). Although this maximises the contamination of the aperture by these nearby stars, it minimises the noise due to the variations in this contamination due to changes in the fraction of the stars' PSFs inside the photometric aperture as the field of view rotates.

  \begin{table*}
    \centering
    \caption{Log of \CHEOPS\ observations.  Effic. is the fraction of the observing interval covered by valid observations of the target. The variables in final column are spacecraft roll angle, $\phi$, and aperture contamination, \var{contam}. } 
    \label{tab:obslog}
    \begin{tabular}{llrrrrl}
\hline
\multicolumn{1}{l}{File key} & 
\multicolumn{1}{l}{Event} & 
\multicolumn{1}{l}{Start date [UTC]} & 
\multicolumn{1}{l}{Duration [s]} & 
\multicolumn{1}{l}{$N_{\rm obs}$} & 
\multicolumn{1}{l}{Effic. [\%]} & 
\multicolumn{1}{l}{Decorrelation parameters} \\ 
\hline
CH\_PR100037\_TG011601\_V0200 & Transit & 2020-11-24T15:41:07 & 48\,682 & 429 & 52.8 & \var{contam}, $\sin\phi$, $\cos3\phi$ \\
CH\_PR100037\_TG017101\_V0200 & Transit & 2021-10-19T00:20:09 & 48\,983 & 519 & 63.5 &\var{contam}, $\sin\phi$, $\cos\phi$,$\sin2\phi$ \\
CH\_PR100037\_TG017201\_V0200 & Eclipse & 2021-09-28T03:09:09 & 34\,936 & 338 & 57.9 & \var{contam}, $\sin\phi$, $\cos2\phi$,$\sin3\phi$  \\
\noalign{\smallskip}
\hline
\end{tabular}
\end{table*}

\subsection{CFHT SPIRou spectroscopy}
 SPIRou is a fibre-fed, cross-dispersed échelle spectrograph mounted on the Canada-France-Hawaii telescope (CFHT) on Maunakea, Hawaii. The spectrograph provides spectra covering the entire wavelength range from 0.95 to 2.35 microns at a spectral resolving power $R\approx75,000$ \citep{2020MNRAS.498.5684D}. 22 spectra of EBLM~J0113+31 with a signal-to-noise per pixel between 77 and 103 near 1 micron were obtained on separate nights between 2020-02-05 and 2020-08-01.
 
 We used spectra extracted from the raw data provided by the observatory using data reduction system (DRS) version 0.6.131. We dealt with the data order-by-order, selecting only those orders with little contamination due to telluric features. The selected wavelength regions are listed in Table~\ref{tab:orders}. The corrections for the échelle blaze function and telluric absorption provided by the DRS were applied. 

 % SPIRou orders table
\begin{table}
	\centering
	\caption{SPIRou échelle orders used in this analysis. Features typically visible in the spectra of M~dwarfs from \citet{1994MNRAS.267..413J} are listed with wavelengths in nm in the final column.}
	\label{tab:orders}
	\begin{tabular}{lrrl} % four columns, alignment for each
		\hline
		Order & $\lambda_{\rm min}$ [nm] & $\lambda_{\rm max}$ [nm] & Notes \\
		\hline
32 &  2363 &  2437 & \\
33 &  2291 &  2363 & \\
34 &  2224 &  2294 & \\
35 &  2160 &  2228 & Na\,I 2206, 2209 \\
36 &  2100 &  2166 & \\
37 &  2043 &  2108 & \\
44 &  1718 &  1772 & \\
45 &  1680 &  1733 & \\
46 &  1643 &  1695 & Al 1676, 1677 \\
47 &  1608 &  1659 & \\
48 &  1575 &  1624 & \\
58 &  1303 &  1344 & \\
59 &  1281 &  1321 & Ca\,I 1313 \\
62 &  1219 &  1257 & K\,I 1243, 1252 \\
63 &  1199 &  1237 & \\
64 &  1181 &  1218 & VO 1200 \\
65 &  1162 &  1199 & K\,I 1169, 1177, 1178 \\
66 &  1145 &  1181 & \\
72 &  1049 &  1082 & \\
73 &  1035 &  1067 & \\
74 &  1021 &  1053 & \\
75 &  1007 &  1039 & \\
78 &   968 &   999 & FeH 990 \\
79 &   956 &   986 & \\
\hline
	\end{tabular}
\end{table}

\subsection{TESS photometry}
 
\begin{figure}
	\includegraphics[width=\columnwidth]{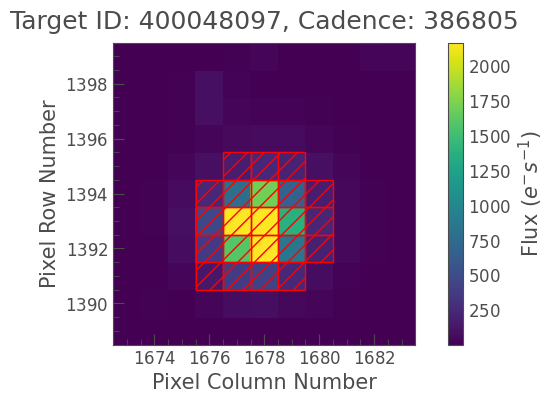}
    \caption{A typical image of EBLM~J0113+31 from the TESS target pixel file showing the pixels used to extract the light curve (red hatching).}
    \label{fig:lightkurve_aperture}
\end{figure}

One transit and two secondary eclipses of EBLM~J0113+31 were observed at 120\,s cadence by the Transiting Exoplanet Survey Satellite \citep[TESS,][]{2015JATIS...1a4003R} in Sector 17 of the primary mission. The TESS target pixel files were downloaded from the Mikulski Archive for Space Telescopes\footnote{\url{https://archive.stsci.edu/}} (MAST) and processed to produce a light curve using the package \software{Lightkurve 2.0} \citep{2018ascl.soft12013L}.\footnote{\url{https://docs.lightkurve.org/}} The pixels used to extract the photometry from the target pixel file are shown in Fig.~\ref{fig:lightkurve_aperture}. Instrumental noise was removed using the cotrending basis vectors (CBVs) provided by the TESS Science Processing Operations Center (SPOC) \citep{2016SPIE.9913E..3EJ}. 
We used 16 ``Single-Scale'' and 7 ``Spike'' CBVs to model trends present in all targets on the same CCD as EBLM~J0113+31. The amplitude of each CBV was determined using only data outside the eclipses and transit. We set the  L2-norm penalty  to $\alpha=0.1$ to achieve a balance between over-fitting the data and effectively removing instrumental trends. 

\section{Analysis}

\subsection{Radial velocity measurements from the SPIRou data}
 We use synthetic spectra taken from \citet{2013A&A...553A...6H}\footnote{\url{http://phoenix.astro.physik.uni-goettingen.de/}} to produce a template for the spectrum of the G0V primary star, using linear interpolation to create a spectrum appropriate for $T_{\rm eff}=6150$\,K, ${\rm [Fe/H]} = -0.4$, $\log g =  4.15$ and $[\alpha/{\rm Fe}] = 0.0$. We then measure the position of the peak in the cross-correlation against this template for the observed spectra order-by-order. Low-frequency noise in the data for each order was removed prior to cross-correlation using a 5-th order high-pass Butterworth filter with a critical frequency of 16/4096 pixels$^{-1}$. We then reject measurements more than 5\,km\,s$^{-1}$ from the median before calculating the mean and standard error in the mean given in Table~\ref{table:rv}. 
  
\begin{table}
\centering
\caption{Radial velocity measurements for EBLM~J0113+31\,A measured from the SPIRou spectra of EBLM~J0113+31. The number of orders used to calculate the mean and standard error on the mean is given in the final column.}
\label{table:rv}
\begin{tabular}{lrrr} % four columns, alignment for each
\hline
Exposure number & \multicolumn{1}{l}{BJD$_{\rm TDB}$}& \multicolumn{1}{c} {$V_r$ [km/s]} & $N$ \\
\hline
2503696 & 2459063.1144 & $    16.72 \pm  0.14$ & 24 \\
2502923 & 2459059.1322 & $    27.01 \pm  0.14$ & 24 \\
2469680 & 2458896.7054 & $    -1.79 \pm  0.18$ & 24 \\
2499300 & 2459038.1105 & $     3.12 \pm  0.16$ & 24 \\
2498079 & 2459033.1244 & $    22.56 \pm  0.11$ & 24 \\
2493617 & 2459011.1184 & $    -1.84 \pm  0.13$ & 24 \\
2498553 & 2459035.1050 & $    14.54 \pm  0.13$ & 23 \\
2499489 & 2459039.0948 & $    -0.06 \pm  0.08$ & 24 \\
2468747 & 2458885.7312 & $     5.73 \pm  0.15$ & 22 \\
2502578 & 2459056.1340 & $    -1.91 \pm  0.13$ & 24 \\
2469510 & 2458895.7054 & $     2.27 \pm  0.16$ & 23 \\
2497677 & 2459031.1037 & $    27.73 \pm  0.12$ & 24 \\
2498357 & 2459034.1213 & $    18.36 \pm  0.16$ & 23 \\
2499115 & 2459037.1309 & $     6.50 \pm  0.17$ & 24 \\
2499760 & 2459040.1279 & $    -2.89 \pm  0.16$ & 23 \\
2498783 & 2459036.1298 & $    10.66 \pm  0.15$ & 23 \\
2468572 & 2458884.7115 & $    -2.25 \pm  0.18$ & 24 \\
2497879 & 2459032.1204 & $    25.75 \pm  0.14$ & 24 \\
2499953 & 2459041.1116 & $    -3.52 \pm  0.13$ & 23 \\
2503110 & 2459060.0927 & $    27.40 \pm  0.15$ & 23 \\
2469883 & 2458897.7056 & $    -3.43 \pm  0.14$ & 23 \\
2470074 & 2458898.7057 & $    -3.21 \pm  0.12$ & 24 \\
\hline
	\end{tabular}
\end{table}
%%%%%%%%%%%%%%%%%%%%%%%%%%%%%%%%%%%%%%%%%%%%%%%%%%

\subsection{Pre-processing of the SPIRou data}
 
 The M~dwarf contributes less than 2\,per~cent of the flux at 1\,$\mu$m so we removed the spectral features in the SPIRou data due to the G0V primary star prior to our attempt to detect the faint companion in these spectra. We use the spectroscopic orbit from GMC+2014 to shift the template spectrum for the primary star to the radial velocity corresponding to the time of mid-exposure for each SPIRou spectrum and then divide the observed spectrum by the shifted model spectrum. 
 
The correction for telluric absorption in the observed spectra will be imperfect so we mask pixels where the telluric absorption is greater than 50\,per~cent. We also mask all pixels in order 47 at wavelengths $>1616$\,nm because there is a strong telluric absorption band at these wavelengths. The removal of spectral features from the primary star will also be imperfect so we mask pixels where absorption lines in the template spectrum are deeper than 50\,per~cent. We then flatten the spectrum by dividing the data by a 16th-order polynomial fit by least-squares to the unmasked data in each order. 
  
Outliers due to cosmic ray hits on the detector and other image anomalies were then identified and removed by flagging pixels more than 4 times the inter-quartile range from the mean in 10 blocks of data per order.

The signal-to-noise is similar for each spectrum but varies quite strongly with wavelength so we use 1.25$\times$ the mean absolute deviation of the data across the observed spectra to assign a standard error to the pixels at each wavelength.

\subsection{Detection of the M dwarf in the SPIRou spectra}
\label{sec:ccfstack}
 The signal from the M~dwarf is too weak to be detected in the individual SPIRou spectra, but it is possible to measure the semi-amplitude of M~dwarf's spectroscopic orbit, $K_2$, by calculating the average cross-correlation function against a suitable template spectrum after shifting these CCFs to the rest frame of the M~dwarf assuming a range of $K_2$ values. The barycentric radial velocity of the M~dwarf at the time of mid-exposure for each spectrum is
 \begin{equation}
 \label{eq:K2}
 V_{r,2} = K_2\,[\cos(\nu+\omega_2)+e\cos(\omega_2)]
 \end{equation}
 The value of the eccentricity $e$ and the longitude of periastron $\omega_2 = \omega_1+\pi$ are known accurately from the spectroscopic orbit of the primary star with longitude of periastron  $\omega_1$, taken from GMC+2014. Similarly, the true anomaly at the time of mid-exposure, $\nu$, can be accurately predicted from the values of $T_0$ (time of mid-transit), $P$ (orbital period),  $e$ and $\omega_1$, also taken from GMC+2014.
 
\begin{figure}
	\includegraphics[width=\columnwidth]{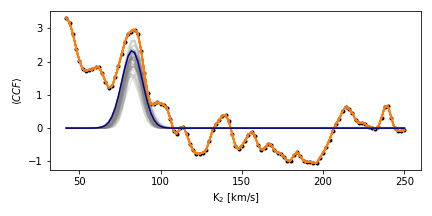}
	\includegraphics[width=\columnwidth]{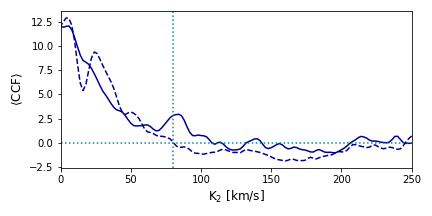}
	\includegraphics[width=\columnwidth]{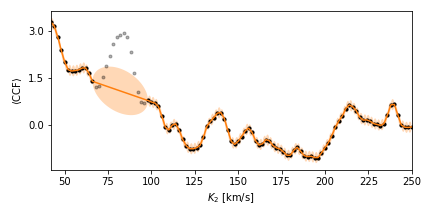}
	\includegraphics[width=\columnwidth]{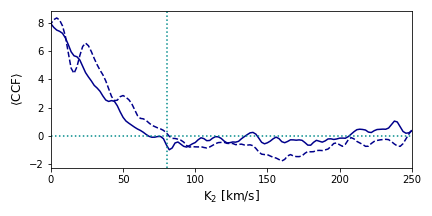}
    \caption{Mean cross-correlation function of EBLM~J0113+31 after shifting to the rest frame of EBLM~J0113+31\,B assuming a range of $K_2$ values. {\it Upper panel:} Gaussian process fit (orange) of a Gaussian profile to the peak near $K_2=83$\,km/s in the stacked CCF (black points). The maximum-likelihood Gaussian profile is plotted in dark blue and 50 samples from the posterior probability distribution are plotted in light grey. {\it Upper middle panel:} the stacked CCF (solid line) and its reflection about $K_2=0$ (dashed line). The  estimated value of $K_2=80.3$\,km/s from GMC+2014 is indicated by a vertical dotted line. {\it Lower middle panel:} Gaussian process fit to the stacked CCF excluding the peak near $K_2=83$\,km/s. The orange shaded region  shows $2\times$ the standard error range on the predicted values of the Gaussian process. {\it Lower panel:} the stacked CCF (solid line) and its reflection about $K_2=0$ (dashed line) computed for spectra with the signature of the M~dwarf removed using a model spectrum with $T_{\rm eff} = 3300$\,K.}
    \label{fig:CCF_figure}
\end{figure}

We use synthetic spectra taken from \citet{2013A&A...553A...6H} as a template for the spectra of the M dwarf, using linear interpolation to create a spectrum appropriate for $T_{\rm eff}=3300$\,K, ${\rm [Fe/H]} = -0.4$, $\log g =  5.0$ and $[\alpha/{\rm Fe}] = 0.0$. The cross-correlation function is calculated order-by-order. Low-frequency noise in the data for each order was removed prior to cross-correlation using a 5-th order high-pass Butterworth filter with a critical frequency of 32/4096 pixels$^{-1}$. The data are apodized using a Gaussian filter with a standard deviation of 64 pixels applied to the data at each end of the order. The correlation coefficient for each order is calculated after shifting the template according to radial velocity computed with equation (\ref{eq:K2}) includes the weights calculated from the estimated standard errors on each pixel.  This is repeated for a uniform grid of $K_2$ values. The average CCF over all orders and all exposures as a function of $K_2$ (``stacked CCF'') is shown in Fig.~\ref{fig:CCF_figure}. \citet{2014A+A...572A..50G} estimate that $K_2 = 80.3 \pm 1.5$\,km\,s$^{-1}$. There is indeed a peak in the stacked CCF near this value of $K_2$. To measure the position of this peak we model the stacked CCF as a Gaussian process (GP) plus a Gaussian profile. We use the \software{celerite} package \citep{celerite1} to compute the likelihood for a GP with a kernel of the form $k(\tau) = a_j\,e^{-c_j\,\tau}$ and the affine-invariant Markov chain Monte Carlo sampler \software{emcee} \citep{2013PASP..125..306F} to sample the posterior probability distribution for the model parameters. Based on this analysis, the peak in the stacked CCF occurs at $K_2 = 82.9 \pm 0.7$\,km\,s$^{-1}$ and has a width of $5.7\pm0.6$\,km\,s$^{-1}$. 

The broad peak in the stacked CCF around $K_2=0$ is due to imperfect removal of telluric features and spectral features from the G0V primary star. We compared the stacked CCF to the average CCF computed with negative values of $K_2$ plotted against $|K_2|$, i.e. the mirror image of the stacked CCF. As can be seen in Fig.~\ref{fig:CCF_figure}, there is no corresponding peak at $K_2 \approx -83$\,km\,s$^{-1}$. This reassures us that the peak at $K_2 \approx +83$\,km\,s$^{-1}$ is unlikely to be due to imperfect removal of telluric features or spectral features from the G0V primary star. We used a fit to the stacked CCF done in the same way as above but excluding data around the peak at $K_2 = 83$\,km\,s$^{-1}$ to estimate the statistical significance of this feature. Based on the GP prediction of the correlated noise in this region shown in Fig.~\ref{fig:CCF_figure}, we estimate that the peak height corresponds to a detection with a significance $\approx 4$-$\sigma$. We also verified that the height of the peak in the stacked CCF is very close to the height expected for an M-dwarf companion given the flux ratio $\ell_T\approx 0.00155$ inferred from the depth of the secondary eclipse in the TESS light curve. We subtracted a scaled version of the template M-dwarf spectrum  from the spectra used to compute the stacked CCF based on this flux ratio in the TESS band and re-computed the stacked CCF. As can be seen in Fig.~\ref{fig:CCF_figure}, the resulting stacked CCF has no peak near $K_2 = 83$\,km\,s$^{-1}$. Based on these three tests, we are confident that our detection of the M-dwarf is robust and that the measurement of $K_2$ is reliable.

\subsection{Initial assessment of the CHEOPS data}

We used the software package \software{pycheops} \citep{pycheops} to make an initial assessment of the light curve data from each of the three CHEOPS visits to the target listed in Table~\ref{tab:obslog}. We excluded data from the analysis where the background level in the images due to scattered light is more than 20\,per~cent larger than the median value during the visit. The data file provided by the data reduction pipeline \citep[DRP,][]{2020A&A...635A..24H} includes a quantity {\tt LC\_CONTAM} that is an estimate of the contamination of the photometric aperture by nearby stars. This quantity varies during the orbit because of the rotation of the spacecraft and the strongly asymmetric point spread function of the instrument. This calculation of {\tt LC\_CONTAM} is based on a simulation of the field of view using the mean G-band magnitudes of the target and nearby stars from Gaia DR2. Fig.~\ref{fig:fov} shows the results of this simulation for one image. This contamination estimate does not account for variability of target itself, so we added a new function to \software{pycheops} version 1.0.2 that corrects the measured flux ({\tt FLUX}) by subtracting the value ${\tt LC\_CONTAM}\times10^{-0.4(G-G_0)}$ from {\tt FLUX}. The zero-point $G_0$ is calculated from the average value of  $$-2.5\log[({\tt LC\_CONTAM}+1) \times 10^{-0.4 G}\times f_{\rm frac}/{\tt FLUX} ],$$  where $f_{\rm frac}$ is the fraction of the total flux from the target in the photometric aperture, $G$ is the mean G-band magnitude of the target, and the average is taken over data points outside of transit and eclipse.

\begin{figure}
	\includegraphics[width=\columnwidth]{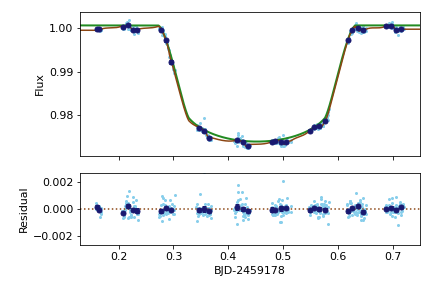}
    \caption{CHEOPS light curve from one visit to observe the transit of EBLM~J0113+31. {\it Upper panel:} The observed light curve is displayed in cyan. The dark blue points are the data points binned over 0.01 phase units. The full model including instrumental trends is shown in brown and the transit model without trends is shown in green.  \textit{Lower panel}: Residuals obtained after subtraction of the best-fit model.} 
    \label{fig:dataset}
\end{figure}

Based on the simulated image of the field of view shown in Fig.~\ref{fig:fov}, we decided to use the OPTIMAL aperture with a radius of 40 pixels for our analysis. This maximises the contamination of the photometric aperture but minimises the uncertainty in this quantity due to errors in measuring the positions of the stars in the image and, hence, the fraction of the flux from each star that is contained in the aperture. We repeated our analysis using the DEFAULT aperture with a radius of 25 pixels and found that the results are entirely consistent with those presented here.  
For each visit we calculate a best-fit for a transit or eclipse model to the light curve including linear decorrelation against {\tt LC\_CONTAM} to account for small errors in estimating the amplitude of the variations in this quantity. We then calculate the best-fit light curves including each of the other available decorrelation parameters and add them one-by-one if the Bayes factor for the parameter exceeds 1. The decorrelation parameters selected by this method are listed in Table~\ref{tab:obslog}. The fit to the data from a typical visit including these detrending parameters in shown in Fig.~\ref{fig:dataset}.

\subsection{Updated transit time ephemeris}
 The two times of mid-transit measured from the CHEOPS data during the initial assessment of the data described above are listed in Table~\ref{tab:tmin} together with the time of mid-transit from GMC+2014 and one new time of mid-transit from a least-squares fit to the TESS light curve  using the transit model from \software{pycheops}. From a least-squares fit to these data we obtain the following updated ephemeris for the times of mid-transit in EBLM~J0113+31:
\label{sec:ephem}
\begin{equation}
\label{eq:ephem}
{\rm BJD} ~T_{\rm mid}  =    2459107.068051(45) +  14.27684012(73)\,{\rm E}
\end{equation}
There is no evidence for any change in orbital period greater than $|\dot{P}/P| \approx 1\times10^{-5}$ from these times of mid-transit.

 \begin{table}
    \centering
    \caption{Times of mid-transit for EBLM~J0113+31. Residuals from the linear ephemeris given in section\ref{sec:ephem} are given in the second column.} 
    \label{tab:tmin}
    \begin{tabular}{rrl}
\hline
\multicolumn{1}{l}{BJD$-2450000$} & $({\rm O}-{\rm C})$ [s]&  Source \\
\noalign{\smallskip}
\hline
$6023.27063 \pm 0.00036 $& $3.9$ & GMC+2014 \\
$8778.70047 \pm 0.00042 $& $-22.3$ & TESS \\
$9178.45224 \pm 0.00017 $& $-1.0$ & CHEOPS \\
$9506.81960 \pm 0.00013 $& $2.2$ & CHEOPS \\
\noalign{\smallskip}
\hline
\end{tabular}
\end{table}

\subsection{Combined analysis of light curve and radial velocity data}

We used the light curve model {\tt ellc} \citep{2016A&A...591A.111M} to calculate synthetic light curves in the TESS and CHEOPS bands, and the spectroscopic orbit of the primary star. This model gives us more flexibility in choosing the level of numerical noise in these synthetic light curves than is possible with the qpower2 algorithm used in \software{pycheops} \citep{2019A+A...622A..33M}. For the analysis presented here we used the ``default'' grid size for the primary star and the ``very\_sparse'' grid size for the companion, which gives numerical noise of only a few ppm at most orbital phases and everywhere less than 10\,ppm. We also tested for the impact of the gravitational distortion of the stars by their mutual gravity on the light curve. This is less than 1.5\,ppm through the transit so we assumed spherical stars for our analysis in order to speed-up the calculation.

The parameters of the binary star model are: the radii of the stars in units of the semi-major axis (fractional radii), $r_1=R_1/a$ and $r_2 = R_2/a$; the surface brightness ratios in the TESS and CHEOPS bands, $S_{\rm T}$ and  $S_{\rm C}$, respectively; the orbital inclination, $i$; the time of mid-transit, $T_0$; the orbital period, $P$;  $f_s = \sqrt{e}\sin(\omega)$ and $f_c = \sqrt{e}\cos(\omega)$, where $e$ is the orbital eccentricity and $\omega$ is the longitude of periastron for the primary star; the semi-amplitude of the primary star's spectroscopic orbit, $K_1$; the limb-darkening parameters assuming a power-2 limb-darkening law, $h_{\rm 1,T}$ and $h_{\rm 2,T}$ in the TESS band, and $h_{\rm 1,C}$ and $h_{\rm 2,C}$ in the CHEOPS band. The ephemeris for the time of mid-transit derived in Section~\ref{sec:ephem} is very accurate so we fix $T_0$ and $P$ at these values in our analysis. The curvature of the light curve between the second and third contact points is very clearly seen in the CHEOPS and TESS light curves, and is almost directly related to the parameters $h_{\rm 1,C}$ and $h_{\rm 1,T}$, respectively,  so we leave this as a free parameter in the analysis. The parameters $h_{\rm 2,C}$ and $h_{\rm 2,T}$ will have a much more subtle influence on the light curve that is almost entirely confined to the ingress and egress phases so we impose priors on these parameters based on the tabulated values of $h_2$ in the TESS and CHEOPS bands from \citet{2018A+A...616A..39M}. The width of the priors is based on the comparison of these tabulated values to the observed values of this parameter from an analysis of the {\it Kepler} light curves of transiting exoplanet systems in the same study.

Prior to the analysis of the CHEOPS data combined with the other data sets, we applied a correction for hot pixels in the photometric aperture. Quantitatively, we define hot pixels as pixels with dark current above 3~e$^-$\,s$^{-1}$. Since the beginning of the mission, hot pixels have appeared regularly in the CHEOPS CCD at a rate of $\sim$100 new hot pixels per day. The CHEOPS Instrument Team monitors closely the number and location of hot pixels. Approximately once per week, ``dark images'' are acquired for that purpose (10 full frame images obtained observing a region of the sky void of stars). These images are used to produce the reference files that track the location and dark current of hot pixels. These reference file are available from the CHEOPS data archive.\footnote{\url{https://cheops-archive.astro.unige.ch/archive_browser/}} We used hot pixel maps generated about 2 days after each visit to EBLM~J0113+31 to calculate the contribution of these hot pixels to the count rate in the photometric aperture. The hot-pixel contamination is $\approx 0.6$\,per~cent in the OPTIMAL aperture for the visit in 2020 and $\approx 1.2$\,per~cent for the visits in 2021. The hot-pixel contamination in the DEFAULT aperture is $\approx 0.3$\,per~cent for all visits. The hot pixel contamination is calculated for every image but the variation in this quantity is small ($\loa 10$\,per~cent of its value) so we apply the correction by subtracting the mean value of the contamination during the visit from the count rate. 

Our model includes the parameter $\ell_{\rm 3,C}$ that is a constant added to the synthetic CHEOPS light curve to account for contamination of the photometric aperture. We applied a correction to the light curves for contamination prior to the combined analysis so, to account for uncertainties in these corrections, we assign a Gaussian prior to $\ell_{\rm 3,C}$ with a mean value of 0 and a standard error equal to 50\,per~cent of the total contamination estimate. Similarly, the parameter $\ell_{\rm 3,T}$ accounts for the contamination of the photometric aperture shown in Fig.~\ref{fig:lightkurve_aperture} used to extract the TESS light curve. We noticed that the entry TIC~400048098 in the TESS input catalogue \citep[TIC,][]{2019AJ....158..138S}  has no counterpart in GAIA EDR3 \citep{2016A&A...595A...1G, 2021A&A...649A...1G} so we assume that this is a spurious entry and do not include it in our calculation of the contamination. The star TIC~400048094 appears near the edge of the default photometric aperture provided with the target pixel file. We added one pixel to this aperture so that there is no ambiguity over whether this star should be included in the calculation of the contamination or not. From the $T$ magnitudes listed the TIC we estimate $\ell_{\rm 3,T} = 0.0030$. We allow this parameter to vary in the fit but assign a Gaussian prior to it, assuming an arbitrary uncertainty of 50\,per~cent.

 The light curves produced by CHEOPS are known to have very low levels of instrumental noise after decorrelation. Similarly, the TESS light curve following correction for instrumental trends that we calculated with \software{lightkurve} shows little sign of residual instrumental noise or stellar variability. We therefore adopt a white noise model for our analysis and assume that the standard deviation per point in the TESS and CHEOPS light curves -- $\sigma_{\rm T}$ and  $\sigma_{\rm C}$, respectively -- are the same for all data points from the same instrument. The logarithm of these standard errors are included as a hyperparameters in our analysis by correctly normalising the calculation of the posterior probability distribution. We only include data from the TESS light curves at orbital phases near the transit and eclipses in this analysis. For both the CHEOPS and TESS data, each section of data around a transit or eclipse is divided by a straight line fit to the data either side of the transit or eclipse prior to analysis. 
 
 We use all the radial velocities published by GMC+2014 plus the new radial velocities from Table~\ref{table:rv} in our analysis. We see no evidence for excess noise in the radial velocities so we use their standard errors as quoted for the calculation of the posterior probability distribution.
 
 In total, we are using nine sets of data, each of which has an uncertain zero-point that should be included in the analysis. Additionally, there are eleven basis functions that are used for the removal of instrumental noise from the CHEOPS data, each with its own coefficient that should be varied independently during the fit to the data. To avoid explicitly calculating these nuisance parameters we use the procedure described by \citet{2017RNAAS...1....7L}, in which the likelihood for any proposed set of model parameters marginalised over a set of nuisance parameters for a linear model can be calculated by modifying the covariance matrix.
 
\begin{figure}
	\includegraphics[width=\columnwidth]{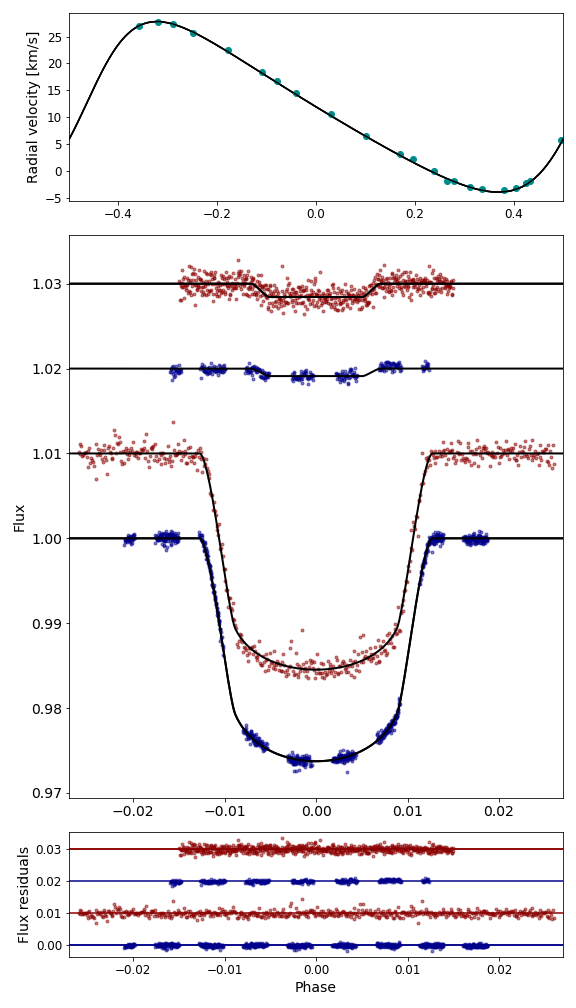}
    \caption{{\it Top panel:} radial velocity measurements for EBLM~J0113+31 (points) and our maximum-likelihood model (line) based on a fit to the combined radial-velocity and light-curve data. Middle panel: TESS (red points) and CHEOPS (blue points) photometry of the transit and eclipse in EBLM~J0113+31. The maximum-likelihood models based on a fit to the combined  radial-velocity and light-curve data is also shown (lines). Data obtained during around the eclipses are plotted as function of orbital phase $-0.5323$. Data and models have been offset vertically for clarity. The CHEOPS data have been corrected for instrumental noise calculated as part of the analysis. {\it Lower panel:} Residuals from the maximum likelihood models plotted in the middle panel.}
    \label{fig:lcrvfit}
\end{figure}

We used {\sc emcee} \citep{2013PASP..125..306F}, a {\sc python} implementation
of an affine invariant Markov chain Monte Carlo (MCMC) ensemble sampler, to
calculate the posterior probability distribution of the model parameters.  The maximum-likelihood model fit to the data is shown in Fig.~\ref{fig:lcrvfit}. The mean and standard error of the posterior probability distributions for each of the model parameters and various derived parameters are given in Table~\ref{tab:lcrvfit}.

The parameters in Table~\ref{tab:lcrvfit} can be combined with our measured value of $K_2$ from the analysis of the SPIRou spectra to determine the masses and radii of both stars with no additional model input. To account for the correlations between parameters, we do this using the sampled posterior probability distribution for the relevant parameters generated by \software{emcee} together with a sample of $K_2$ values assuming a Gaussian distribution for this parameter. The masses and radii derived are given in Table~\ref{tab:mr}.

 \begin{table}
    \centering
    \caption{Fit to RV and LC data. ${\cal N}(\mu, \sigma)$ denotes a Gaussian prior applied to a parameter with mean $\mu$ and standard deviation $\sigma$.} 
    \label{tab:lcrvfit}
    \begin{tabular}{lrl}
\hline
\multicolumn{1}{l}{Parameter} &  Value & Notes \\
\hline
\noalign{\smallskip}
 $  R_1/a              $&$  0.05348 \pm  0.00031 $& \\
 $  R_2/a              $&$ 0.008111 \pm 0.000063 $& \\
 $  i\: [^{\circ}]     $&$   89.110 \pm    0.041 $& \\ % Inclination\\
 $  f_s                $&$ -0.54885 \pm  0.00043 $& \\ % $\sqrt{e}\sin\omega$ \\
 $  f_c                $&$  0.08693 \pm  0.00026 $& \\ % $\sqrt{e}\cos\omega$ \\
 $  S_{\rm T}          $&$   0.0675 \pm   0.0033 $& \\ % Surface brightness ratio, TESS \\
 $  S_{\rm C}          $&$   0.0384 \pm   0.0023 $& \\ % Surface brightness ratio, CHEOPS \\
 $  K_1     $ [km/s]    &$   15.861 \pm    0.010 $& \\
 $  h_{\rm 1,C}        $&$   0.7683 \pm   0.0038 $& \\
 $  h_{\rm 2,C}        $&$    0.720 \pm    0.036 $& ${\cal N}(0.409, 0.045)$ \\
 $  h_{\rm 1,T}        $&$   0.8008 \pm   0.0074 $& \\
 $  h_{\rm 2,T}        $&$    0.779 \pm    0.022 $& ${\cal N}(0.379, 0.045)$ \\
 $ \ell_{\rm 3,C}      $&$    0.007 \pm    0.009 $& ${\cal N}(0.000, 0.012)$, 1 \\
 $ \ell_{\rm 3,T}      $&$    0.019 \pm    0.010 $& ${\cal N}(0.030, 0.015)$, 2 \\
 $  \ln \sigma_{\rm C} $&$    -7.80 \pm     0.02 $& \\
 $  \ln \sigma_{\rm T} $&$    -7.04 \pm     0.02 $& \\
\noalign{\smallskip}
\multicolumn{3}{l}{Derived parameters} \\
 $ e                          $&$      0.30879 \pm  0.00045 $&  \\
 $ \omega\: [^{\circ}]        $&$      279.000 \pm    0.031 $&  \\
 $ \sin i                     $&$      0.99988 \pm  0.00001 $&  \\
 $ R_2/R_1                    $&$      0.15164 \pm  0.00073 $&  \\
 $ \ell_{\rm T}               $&$      0.00155 \pm  0.00008 $& Flux ratio, TESS \\
 $ \ell_{\rm C}               $&$      0.00088 \pm  0.00005 $& Flux ratio, CHEOPS \\
 $ \sigma_{\rm T} $ [ppm]      &$          874 \pm       18 $& \\ % Standard error per point, TESS \\
 $ \sigma_{\rm C} $ [ppm]      &$          410 \pm        8 $& \\ % Standard error per point, CHEOPS \\
\noalign{\smallskip}
\hline
\end{tabular}
\parbox{\columnwidth}{
1: After correction for contamination of the photometric aperture by nearby stars and hot pixels. 2: Including flux from other stars in the photometric aperture.}
\end{table}

%__________________________________________________ MASS RADIUS
\begin{table}
\caption[]{Mass, radius, effective temperature and derived parameters for the stars in EBLM~J0113+31. The metallicity [M/H] is estimated from our analysis of the spectrum of EBLM~J0113+31\,A.  N.B. $T_{\rm eff,1}$  and $T_{\rm eff,2}$ are subject to additional systematic uncertainty of 10\,K and 7\,K, respectively.}
\label{tab:mr}
\begin{center}
  \begin{tabular}{lrrr}
\hline
\noalign{\smallskip}
 \multicolumn{1}{l}{Parameter} &
 \multicolumn{1}{l}{Value} &
 \multicolumn{1}{l}{Error} &
 \multicolumn{1}{r}{} \\
\noalign{\smallskip}
\hline
\noalign{\smallskip}
$M_1/{\mathcal M^{\rm N}_{\odot}}$&1.029  & $\pm$ 0.025 & [2.4\%] \\
\noalign{\smallskip}
$M_2/{\mathcal M^{\rm N}_{\odot}}$&0.197 & $\pm$ 0.003 &[1.5\%] \\
\noalign{\smallskip}
$R_1/{\mathcal R^{\rm N}_{\odot}}$&1.417 & $\pm$ 0.014 &[1.0\%] \\
\noalign{\smallskip}
$R_2/{\mathcal R^{\rm N}_{\odot}}$&0.215 & $\pm$ 0.002 &[1.1\%] \\
\noalign{\smallskip}
$T_{\rm eff,1} $ [K]  & $ 6124 $ & $\pm$ 40  & [0.6\%] \\
\noalign{\smallskip}
$T_{\rm eff,2} $ [K]  & $ 3375 $ & $\pm$ 40  & [1.3\%] \\
\noalign{\smallskip}
$\rho_1/{\rho^{\rm N}_{\odot}}$  &0.362 & $\pm$ 0.006 &[1.7\%] \\
\noalign{\smallskip}
$\rho_2/{\rho^{\rm N}_{\odot}}$  & 19.9 & $\pm$ 0.5 &[2.4\%] \\
\noalign{\smallskip}
$\log g_1$ [cgs] & 4.148 & $\pm$ 0.006 & [1.5\%]  \\
\noalign{\smallskip}
$\log g_2$ [cgs] & 5.068 & $\pm$ 0.006 & [1.5\%] \\
\noalign{\smallskip}
$\log L_1/{\mathcal L^{\rm N}_{\odot}}$   & 0.406 & $\pm$ 0.014 & [3.2\%]  \\
\noalign{\smallskip}
$\log L_2/{\mathcal L^{\rm N}_{\odot}}$   & $ -2.267 $& $\pm$ 0.024 & [5.5\%]  \\
\noalign{\smallskip}
[M/H] & $-0.3$ & $\pm$ 0.1  &  \\ % Average of SPIRou and FIES results
\noalign{\smallskip}
\hline
\end{tabular}
\end{center}
\end{table}

\subsection{Direct measurement of the stellar effective temperature}
The effective temperature for a star with Rosseland radius $R$ and total luminosity $L$ is defined by the equation  
$$L=4\pi R^2 \sigma_{\rm SB} {\rm T}_{\rm eff}^4,$$
where $\sigma_{\rm SB}$ is the Stefan-Boltzmann constant. For a binary star at distance $d$, i.e. with parallax $\varpi=1/d$, the flux corrected for extinction observed at the top of Earth's atmosphere is
$$f_{0,b}= f_{0,1}+f_{0,2}=\frac{\sigma_{\rm SB}}{4}\left[\theta_1^2{\rm T}_{\rm eff,1}^4 + \theta_2^2{\rm T}_{\rm eff,2}^4\right],$$
where $\theta_1=2R_1\varpi$ is the angular diameter of star 1, and similarly for star 2. All the quantities are known or can be measured for EBLM~J0113+31 provided we can accurately integrate the observed flux distributions for the two stars independently. This is possible because photometry of the combined flux from both stars is available from ultraviolet to mid-infrared wavelengths, and the flux ratio at wavelengths where the majority of the flux is emitted by the primary star has been measured from the TESS and CHEOPS light curves. Although we have no direct measurement of the flux ratio at infrared wavelengths, we can make a reasonable estimate for the small contribution of the M-dwarf to the measured total infrared flux using empirical colour\,--\,$T_{\rm eff}$ relations. The M-dwarf contributes less than 0.2\,per~cent to the total flux so it is not necessary to make a very accurate estimate of the M-dwarf flux distribution in order to derive an accurate value of $T_{\rm eff}$ for the G0-type primary star.

The photometry used in this analysis is given in Table~\ref{tab:mags}. The NUV and FUV magnitudes are taken from GALEX data release GR7 \citep{2014yCat.2335....0B} with a correction to the IUE flux scale based on the results from \citet{2014MNRAS.438.3111C}. We assume that the flux from the M-dwarf at ultraviolet wavelengths is negligible. The Gaia photometry is from Gaia data release EDR3. J, H and Ks magnitudes are from the 2MASS survey \citep{2006AJ....131.1163S}. WISE magnitudes are from the All-Sky Release Catalog \citep{2012yCat.2311....0C} with corrections to Vega magnitudes made as recommended by \citet{2011ApJ...735..112J}. Photometry in the PanSTARRS-1 photometry system is taken from \citet{2018ApJ...867..105T}. Details of the zero-points and response functions used to calculate synthetic photometry from an assumed spectral energy distribution are given in \citet{2020MNRAS.497.2899M}.

 To estimate the reddening towards EBLM~J0113+31 we use the calibration of E(B$-$V) versus the equivalent width of the interstellar Na\,I~D$_1$ line by \citet{1997A&A...318..269M}. To measure EW(Na\,I~D$_1$) we used 11 spectra obtain with the FIES spectrograph on the Nordic Optical Telescope used in medium resolution mode ($R = 46,000$). We first shifted these spectra into the rest frame of the primary star and then took the median value at each wavelength to obtain a high signal-to-noise spectrum of the G0V primary star. We then divided each observed spectrum by this spectrum of the G0V primary star after shifting it back to the barycentric rest frame. We then took the median of these residual spectra to obtain a high signal-to-noise spectrum of the interstellar features. The equivalent width of the Na\,I~D$_1$ line measured by numerical integration is EW(Na\,I~D$_1) = 77.1 \pm 6.0$\,m\AA. This value is less than the values of EW(Na\,I~D$_1$) for all the stars in the  calibration sample of \citet{1997A&A...318..269M}. To estimate the uncertainty on the value of E(B$-$V) for EBLM~J0113+31 we take the sample standard deviation for the 5 stars in the calibration sample with the lowest values of  EW(Na\,I~D$_1)\approx 250$\,m\AA. Based on this analysis we obtain the estimate E(B$-$V$)= 0.002 \pm 0.012$. We use this as a Gaussian prior in our analysis but exclude negative values of E(B$-$V).  %The co-added FIES spectra can be obtained from the supplementary online information that accompanies this article.

To establish colour\,--\,$T_{\rm eff}$ relations suitable for dwarf stars with $3100\,{\rm K} < {\rm T}_{\rm eff} < 3500\,{\rm K} $ we use a robust linear fit to the stars listed in Table~6  of \citet{2018MNRAS.475.1960F} within this $T_{\rm eff}$ range. Photometry for these stars is taken from the TESS input catalogue. To estimate a suitable standard error for a Gaussian prior based on this fit we use 1.25$\times$ the mean absolute deviation of the residuals from the fit. Colour\,--\,$T_{\rm eff}$ relations suitable for the primary G0V star were calculated in similar way based on stars selected from the Geneva-Copenhagen survey \citep{2009A&A...501..941H, 2011A&A...530A.138C}  with $5950\,{\rm K} < {\rm T}_{\rm eff} < 6250\,{\rm K}$, $E({\rm B}-{\rm V})<0.01$ and $3.5 < \log g < 4.5$. The results are given in Table~\ref{tab:frp}. 

\begin{table*}
\centering
\caption{Observed apparent magnitudes for EBLM~J0113+31 and predicted values based on our synthetic photometry. The predicted magnitudes are shown with error estimates from the uncertainty on the zero-points for each photometric system. The pivot wavelength for each band pass is shown in the column headed $\lambda_{\rm pivot}$. The magnitudes of the primary G0V star alone corrected for the contribution to the total flux from the M-dwarf are shown in the column headed $m_1$. The flux ratio in each band is shown in the fina column.}
\label{tab:mags}
\begin{tabular}{lrrrrrr} % four columns, alignment for each
\hline
		Band &  $\lambda_{\rm pivot}$ [nm]& \multicolumn{1}{c}{Observed} & \multicolumn{1}{c}{Computed} & 
\multicolumn{1}{c}{$\rm O-\rm C$} & \multicolumn{1}{c}{$m_1$} & \multicolumn{1}{c}{$\ell$} [\%]\\
\hline
FUV           &   154 & $  20.01 \pm  0.54$ & $ 20.74 \pm  0.13$ & $ -0.73 \pm  0.55$ & $  20.01 \pm   0.54$ &  0.00 \\
NUV           &   230 & $  14.28 \pm  0.71$ & $ 14.41 \pm  0.15$ & $ -0.13 \pm  0.73$ & $  14.28 \pm   0.71$ &  0.00 \\
G             &   622 & $  9.920 \pm 0.003$ & $ 9.919 \pm 0.003$ & $+0.002 \pm 0.004$ & $  9.922 \pm  0.003$ &  0.09 \\
BP            &   511 & $ 10.197 \pm 0.003$ & $10.202 \pm 0.003$ & $-0.005 \pm 0.004$ & $ 10.197 \pm  0.003$ &  0.04 \\
RP            &   777 & $  9.477 \pm 0.004$ & $ 9.475 \pm 0.004$ & $+0.002 \pm 0.005$ & $  9.479 \pm  0.004$ &  0.17 \\
g$_{\rm P1}$  &   485 & $ 10.249 \pm 0.020$ & $10.234 \pm 0.005$ & $+0.015 \pm 0.021$ & $ 10.249 \pm  0.020$ &  0.03 \\
r$_{\rm P1}$  &   620 & $  9.961 \pm 0.024$ & $ 9.911 \pm 0.005$ & $+0.050 \pm 0.025$ & $  9.962 \pm  0.024$ &  0.07 \\
i$_{\rm P1}$  &   754 & $  9.868 \pm 0.021$ & $ 9.820 \pm 0.005$ & $+0.048 \pm 0.022$ & $  9.870 \pm  0.021$ &  0.16 \\
J             &  1241 & $  8.982 \pm 0.024$ & $ 8.973 \pm 0.005$ & $+0.009 \pm 0.025$ & $  8.987 \pm  0.024$ &  0.45 \\
H             &  1650 & $  8.692 \pm 0.029$ & $ 8.713 \pm 0.005$ & $-0.021 \pm 0.029$ & $  8.699 \pm  0.029$ &  0.60 \\
K$_{\rm s}$   &  2164 & $  8.620 \pm 0.024$ & $ 8.652 \pm 0.005$ & $-0.032 \pm 0.025$ & $  8.628 \pm  0.024$ &  0.73 \\
W1            &  3368 & $  8.590 \pm 0.023$ & $ 8.613 \pm 0.002$ & $-0.023 \pm 0.023$ & $  8.600 \pm  0.023$ &  0.91 \\
W2            &  4618 & $  8.629 \pm 0.020$ & $ 8.619 \pm 0.002$ & $+0.010 \pm 0.020$ & $  8.642 \pm  0.020$ &  1.18 \\
W3            & 12073 & $  8.633 \pm 0.021$ & $ 8.617 \pm 0.002$ & $+0.016 \pm 0.021$ & $  8.651 \pm  0.021$ &  1.64 \\
W4            & 22194 & $   8.38 \pm  0.22$ & $ 8.674 \pm 0.002$ & $ -0.29 \pm  0.22$ & $   8.40 \pm   0.22$ &  1.64 \\		
\hline
\end{tabular}
\end{table*}

\begin{table}
	\centering
	\caption{Colour-$T_{\rm eff}$ relations used to establish Gaussian priors on the flux ratio at infrared wavelengths for EBLM~J0113+31. The dependent variables are $X_1 = T_{\rm eff,1}-6.1\,{\rm kK}$ and $X_2 = T_{\rm eff,2}-3.3\,{\rm kK}$. }
	\label{tab:frp}
	\begin{tabular}{lll} % four columns, alignment for each
		\hline
		Colour &  Primary & Secondary  \\
		\hline
V$-$J           & $ 1.048 -0.4257 \, X_1 \pm 0.015 $ & $ 4.187 -2.762 \, X_2 \pm  0.11 $ \\
V$-$H           & $ 1.288 -0.5568 \, X_1 \pm 0.019 $ & $ 4.776 -2.552 \, X_2 \pm  0.15 $ \\
V$-$K$_{\rm s}$ & $ 1.357 -0.5926 \, X_1 \pm 0.016 $ & $ 5.049 -2.776 \, X_2 \pm  0.12 $ \\
V$-$W1          & $ 1.405 -0.5829 \, X_1 \pm 0.027 $ & $ 5.207 -2.720 \, X_2 \pm  0.12 $ \\
V$-$W2          & $ 1.411 -0.5753 \, X_1 \pm 0.045 $ & $ 5.365 -2.957 \, X_2 \pm  0.11 $ \\
V$-$W3          & $ 1.355 -0.5919 \, X_1 \pm 0.022 $ & $ 5.477 -3.091 \, X_2 \pm  0.13 $ \\
V$-$W4          & $ 1.397 -0.5812 \, X_1 \pm 0.045 $ & $ 5.620 -3.248 \, X_2 \pm  0.23 $ \\
\hline
\end{tabular}
\end{table}

The method we have developed to measure $T_{\rm eff}$ for eclipsing binary stars is described fully in \citet{2020MNRAS.497.2899M}. Briefly, we use  \software{emcee} \citep{2013PASP..125..306F} to sample the posterior probability distribution $P(\Theta| D)\propto P(D|\Theta)P(\Theta)$ for the model parameters $\Theta$ with prior $P(\Theta)$ given the data, $D$ (observed apparent magnitudes and flux ratios). The model parameters are  $$\Theta = \left({\rm T}_{\rm eff,1},  {\rm T}_{\rm eff,2}, \theta_1, \theta_2, {\rm E}({\rm B}-{\rm V}), \sigma_{\rm ext}, \sigma_{\ell},  c_{1,1}, \dots, c_{2,1}, \dots\right).$$  The prior $P(\Theta)$ is calculated using the angular diameters $\theta_1$ and $\theta_2$ derived from the radii $R_1$ and $R_2$ and the parallax $\varpi$, the priors on the flux ratio at infrared wavelengths based on the colour\,--\,T${\rm eff}$ relations in Table~\ref{tab:frp}, and the Gaussian prior on the reddening described above. The hyperparameters $\sigma_{\rm ext}$ and $\sigma_{\ell}$ account for additional uncertainties in the synthetic magnitudes  and flux ratio, respectively, due to errors in zero-points, inaccurate response functions, stellar variability, etc. The parallax is taken from Gaia EDR3 with corrections to the zero-point from \citet{2022MNRAS.509.4276F}.

To calculate the synthetic photometry for a given value of $T_{\rm eff}$ we used a model spectral energy distribution (SED) multiplied by a distortion function, $\Delta(\lambda)$. The distortion function is a linear superposition of Legendre polynomials in log wavelength. The coefficients of the distortion function for star 1 are $c_{1,1}, c_{1,2}, \dots$, and similarly for star 2. The distorted SED for each star is normalized so that the total apparent flux prior to applying reddening is $\sigma_{\rm SB}\theta^2{\rm T}_{\rm eff}^4/4$. These distorted SEDs provide a convenient function that we can integrate to calculate synthetic photometry that has realistic stellar absorption features, and where the overall shape can be adjusted to match the observed magnitudes from ultraviolet to infrared wavelengths, i.e. {\it the effective temperatures we derive are based on the integrated stellar flux and the star's angular diameter, not SED fitting.}

\begin{figure}
	\includegraphics[width=\columnwidth]{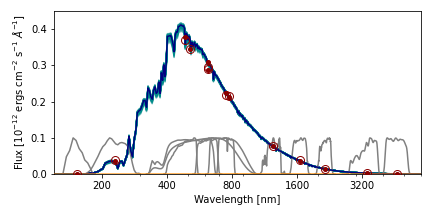}
    \caption{The spectral energy distribution (SED) of EBLM~J0113+31. The observed fluxes are plotted with open circles and the predicted fluxes for the mean of the posterior probability distribution (PPD) integrated over the response functions shown in grey are plotted with filled symbols. The SED predicted by the mean of the PPD is plotted in dark blue and light blue shows the SEDs produced from 100 random samples from the PPD. The contribution to the total SED from the M dwarf (barely visible) is shown in orange. The W3 and W4 mid-infrared bands also used in the analysis are not shown here. }
    \label{fig:sed}
\end{figure}

For this analysis we use model SEDs computed from BT-Settl model atmospheres \citep{2013MSAIS..24..128A} obtained from the Spanish Virtual Observatory.\footnote{\url{http://svo2.cab.inta-csic.es/theory/newov2/index.php?models=bt-settl}} We use linear interpolation to obtain a reference SED for the G0V star appropriate for  $T_{\rm eff,1}=6130$\,K, $\log g_1 = 4.15$, $[{\rm Fe/H}] = -0.3$ and   $[{\rm \alpha/Fe}] = 0.0$. For the reference SED for the M dwarf companion we assume $T_{\rm eff,1}=3380$\,K, $\log g_1 = 5.0$, and the same composition. We experimented with distortion functions with 1, 2, 3, 4 coefficients per star and found the results to be very similar in all cases. The results presented here use two distortion coefficient per star because there is no improvement in the quality if the fit if we use a larger number of coefficients. The predicted apparent  magnitudes including their uncertainties from errors in the zero-points for each photometric system are compared to the observed apparent magnitudes in Table~\ref{tab:mags}. The posterior probability distribution for the model parameters is summarised in Table~\ref{tab:teb} and the spectral energy distribution is plotted in Fig.~\ref{fig:sed}.

The random errors quoted in Table~\ref{tab:teb} do not allow for the systematic error due to the uncertainty in the absolute calibration of the CALSPEC flux scale \citep{2014PASP..126..711B}. This additional systematic error is 10\,K for the G0V primary star and 7\,K for the M-dwarf companion. 

\begin{table}
\centering
\caption{Results from our analysis to obtain the effective temperatures for both stars in EBLM~J0113+31. N.B. $T_{\rm eff,1}$  and $T_{\rm eff,2}$ are subject to additional systematic uncertainty of 10\,K and 7\,K, respectively.}
\label{tab:teb}
\begin{tabular}{lrrl} % four columns, alignment for each
\hline
Parameter & 
\multicolumn{1}{l}{Value} & 
\multicolumn{1}{l}{Error} & Units \\
\hline
\noalign{\smallskip}
$T_{\rm eff,1}$ & $6124 $&$ \pm  40  $ & K \\
$T_{\rm eff,2}$ & $3375 $&$ \pm  40 $ & K \\
$\theta_1$      & $ 0.0745 $&$ \pm 0.0007 $ & mas \\
$\theta_2$      & $ 0.0113 $&$ \pm 0.0001 $ & mas \\
E(B$-$V)        & $ 0.010 $&$ \pm 0.007 $ \\
$\sigma_{\rm ext}$  & $0.014 $&$ \pm 0.011 $ \\
$\sigma_{\ell} $ & $ 0.0002 $&$ \pm 0.0001 $ \\
$c_{1,1}$  & $  0.06  $&$  \pm 0.03 $ \\
$c_{1,2}$  & $ -0.08  $&$  \pm 0.05 $ \\
$c_{2,1}$ & $  0.3  $& $ \pm 0.2  $ \\
$c_{2,2}$ & $ -0.3  $& $ \pm 0.2  $ \\
\hline
\end{tabular}
\end{table}

\subsection{Abundance analysis}

We have used the H-band spectrum of EBLM~J0113+31\,A to estimate this star's metallicity. For this abundance analysis we used the observed SPIRou spectra merged into 1-dimensional spectra provided by the observatory. We first subtracted the model spectrum for the M-dwarf companion described in Section~\ref{sec:ccfstack} from each of these 1-dimensional spectra, scaled such that the flux ratio in the TESS band matched the value measured from the depth of the secondary eclipse in the light curve.  We then co-added the spectra in the rest frame of the primary star to produce a high signal-to-noise spectrum of the G0V star with negligible contamination from the M-dwarf. 
 
\begin{figure*}
	\includegraphics[width=\textwidth]{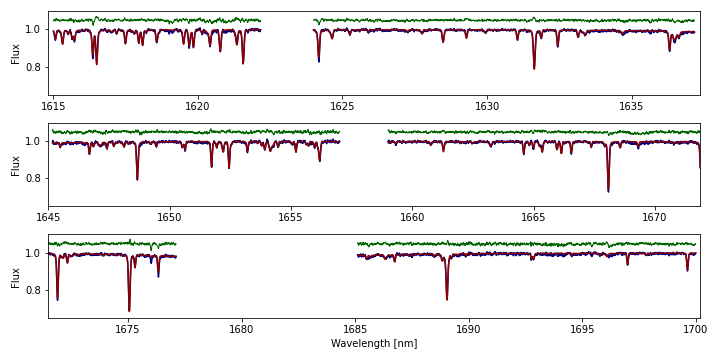}
    \caption{The H-band spectrum of EBLM\,J0113+31\,A (blue) and a synthetic spectrum fit by least-squares using iSpec (red).  Residuals from the synthetic spectrum fit are shown in green offset vertically by 1.05 units.}
    \label{fig:hband}
\end{figure*}

For the analysis of this spectrum we used iSpec \citep{2014A&A...569A.111B, 2019MNRAS.486.2075B} with the APOGEE line list  for atomic and molecular data in the wavelength range 1500\,--\,1700\,nm \citep{2015ApJS..221...24S}. We followed \citet{2020A&A...636A..85S} in selecting Turbospectrum \citep{1998A&A...330.1109A, 2012ascl.soft05004P} for the spectrum synthesis assuming a micro-turbulent velocity $v_{\rm mic} = 1.06$\,km\,s$^{-1}$ with model atmospheres from the MARCS grid \citep{2008A&A...486..951G} and solar abundances from \citet{2007SSRv..130..105G}. We excluded from the fit $\pm 4$\,nm around the two Brackett series lines at 1681.11\,nm and  1641.17\,nm, and also some instrumental features that occur near the ends of the échelle orders at 1657\,--\,1659\,nm and 1622\--\,1624\,nm. We fixed the value of $T_{\rm eff} = 6124$\,K and $\log g = 4.15$. For the macro-turbulent velocity we used the calibration by \citet{2005ApJS..159..141V} to estimate $v_{\rm mac}= 4.67$\,km\,s$^{-1}$. We included the rotational broadening parameter $v\sin i$ as a free parameter in the least-squares fit with a linear limb-darkening coefficient of 0.5 in the H-band based on the results from \citet{2018A&A...618A..20C}.  We attempted a least-squares fit including the $\alpha$-element abundance as a free parameter but found that the value obtain is not accurate enough to be useful so we fixed $[\alpha/{\rm Fe}]=0$ in the least-squares fit. From this least-squares fit we obtained $[{\rm M/H}] = -0.33 \pm 0.01$ and  $v\sin i=6.6\pm0.3$\,km\,s$^{-1}$. There are several additional sources of uncertainty in this analysis, e.g. inaccurate normalisation, errors in atomic data, approximations in the stellar atmosphere models, etc., so the accuracy of our metallicity estimate will be much worse than the precision estimated from the least-squares fitting algorithm \citep{2019MNRAS.486.2075B, 2019ARA&A..57..571J}. Based on the results from independent analyses of APOGEE spectra by \citet{2018AJ....156..126J}, we assume an accuracy of 0.15\,dex, i.e.  $[{\rm M/H}] = -0.33 \pm 0.15$. The fit to the spectrum is shown in Fig.~\ref{fig:hband}. 

We used the co-added FIES spectra of the star to determine the stellar atmospheric parameters ($T_{\mathrm{eff}}$, $\log{g}$, micro-turbulence, and [Fe/H]) and chemical abundances following the methodology described in our previous works \citep[][]{Sousa-14, Santos-13, Adibekyan-12b, Adibekyan-15}. In brief, we make use of the equivalent widths (EW) of spectral lines, as measured using the ARES v2 code\footnote{The last version of ARES code (ARES v2) can be downloaded at \url{http://www.astro.up.pt/~sousasag/ares}}  \citep{Sousa-15}, and we assume ionization and excitation equilibrium. The process makes use of a grid of Kurucz model atmospheres \citep{Kurucz-93} and the radiative transfer code MOOG \citep{Sneden-73}.

For the stellar spectroscopic parameters we obtained $T_{\rm eff} = 6025 \pm 50$\,K, $\log g = 4.10 \pm 0.05$, $V_{\mathrm{tur}} = 1.07 \pm 0.06$\,km\,s$^{-1}$ and $[{\rm Fe/H}] = -0.31 \pm 0.04$. Within the uncertainties, these values are in agreement with those presented in Table|\ref{tab:teb}. In order to be consistent, and because of higher accuracy, we fixed the values of effective temperature and surface gravity to $T_{\mathrm{eff}} = 6124 \pm 40$\,K and $\log g  = 4.148 \pm 0.006$ when determining the abundances of individual elements. Our derivation of three $\alpha$-elements ($[{\rm Mg/H}] = -0.18 \pm 0.09$, $[{\rm Si/H}] = -0.26 \pm 0.04$, $[{\rm Ti/H}] = -0.22 \pm 0.07$) indicates that EBLM J0113+31 is not an $\alpha$-enhanced star ($[\alpha/{\rm Fe}] = 0.09 \pm 0.08$) which is typical for stars in the Galactic thin-disk population \citep{Adibekyan-11}. 

Using the astrometric data from Gaia EDR3 and the radial velocity of the system ($11.179\pm0.004$\,km\,s$^{-1}$, GMC+2014) we calculated the Galactic space velocity components $(U,V,W)  = (-17, 16, 21)$\,km\,s$^{-1}$ with respect to the local standard of rest \citep{Schonrich-10}. Based on these velocities, adopting the characteristics parameters of Galactic stellar populations of \citet{Reddy-06}, and following \citet{Adibekyan-12b} we estimated a probability of 99\% that the star belongs to the Galactic thin disk, which is in agreement with our conclusion based on the composition of the star.

Based on the results from the analysis of the SPIRou and FIES spectra we adopt the value $[{\rm M/H}] = -0.3 \pm 0.1$ for the metallicity of EBLM~J0113+31. The co-added SPIRou spectra corrected for the contribution from the M-dwarf and the co-added FIES spectrum are available from the supplementary online information that accompanies this article.

\section{Discussion}
\label{sec:discuss}

\subsection{Astrometric noise due to binary orbital motion}

The projected semi-major axis of the G0V star's orbit is $\alpha_1 = a_1/d$ = 0.11\,mas, so we expect  excess noise in the Gaia astrometry $\approx 0.1$\,mas due to the orbital motion of the primary star. Indeed, the astrometric excess noise in the Gaia EDR3 catalogue for EBLM~J0113+31 is 0.163 mas. This is higher than expected for a good fit to the data for a single star with ${\rm G}\approx10$, and consistent with the noise expected from the orbital motion of the G0V star. This will only lead to a systematic error in the parallax if the position angle of the binary at the times of observation are not randomly distributed around the binary star orbit. This can be checked using the parameter {\tt ipd\_gof\_harmonic\_amplitude} provided in the EDR3 catalogue \citep{2021A&A...649A...2L}. For EBLM~J0113+31, this parameter takes the value 0.014, which is less than the median value of this statistic for stars with 6-parameter solutions in the magnitude range G=9-12 (0.020). Although the detection of the astrometric noise is statistically significant, it is a small contribution to the uncertainties on the parallax. The renormalised unit weight error for EBLM~J0113+31 is RUWE=1.154, which is only slightly higher than the median value for stars with 6-parameter solutions in the magnitude range G=9-12 (RUWE=1.127), and is close to the expected value $\approx 1$ for ``for well behaved sources''.

We can therefore be confident that the orbital motion of the G0V star does not produce a systematic error in the measured Gaia parallax.

\subsection{Comparison to stellar evolution models}
The mass, radius and effective temperature for both stars in EBLM~J0113+31 are given in Table~\ref{tab:mr}, together with the derived surface gravity, mean stellar density and luminosity for both stars.

\begin{figure}
	\includegraphics[width=0.9\columnwidth]{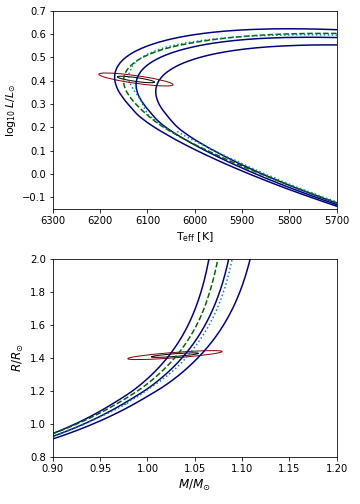}
    \caption{EBLM\,J0113+31\,A in the mass-radius and Hertzsprung-Russell diagrams compared to isochrones for an age of  $6.7\pm0.7$\,Gyr assuming an initial metal abundance $[{\rm Fe/H}]=-0.2$ interpolated from a grid of {\sc garstec} stellar models. The ellipses show 1-$\sigma$ and  2-$\sigma$ confidence regions on the parameters of EBLM\,J0113+31\,A.  Also shown are isochrones for the same age and initial metal abundance from the Dartmouth stellar evolution database (cyan dotted line) and MIST (green dashed line).}
    \label{fig:isochrones1}
\end{figure}

 We used the software package \software{bagemass} \citep{2015A&A...575A..36M} to compare the parameters of the primary star, EBLM~J0113+31\,A, to a grid of stellar models computed with the  {\sc garstec} stellar evolution code \citep{2008Ap&SS.316...99W}. The methods used to calculate the stellar model grid are described in \citet{2013MNRAS.429.3645S}.  \software{bagemass} uses a Markov-chain Monte-Carlo method to explore the posterior probability distribution (PPD) for the mass and age of a star based on its observed $T_{\rm eff}$, luminosity, mean stellar density and surface metal abundance [Fe/H]. We find a very good fit to the observed parameters of EBLM~J0113+31\,A for an age of $6.7\pm0.5$\,Gyr, as can be seen in Fig.~\ref{fig:isochrones1}. More the 99\,per\,cent of samples from the PPD correspond to models where EBLM~J0113+311\,A is a post main-sequence star that has exhausted all the hydrogen in its core. The {\sc garstec} model grid accounts for diffusion so the initial metal abundance for this star is inferred to be ${\rm [Fe/H]} = -0.2 \pm 0.1$. Isochrones for the same age and initial metal abundance from the Dartmouth stellar evolution database \citep[][]{2008ApJS..178...89D} and the MESA Isochrones \& Stellar Tracks \citep[MIST, ][]{2016ApJ...823..102C} are also shown in Fig.~\ref{fig:isochrones1}. There is very good agreement between these different stellar evolution codes, as might be expected given that the properties of EBLM~J0113+31\,A are similar to the Sun and all three grids of stellar models are calibrated to match the observed properties of the Sun.

\begin{figure}
	\includegraphics[width=0.9\columnwidth]{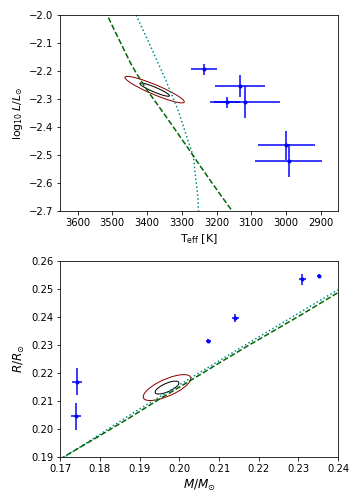}
    \caption{EBLM\,J0113+31\,B in the mass-radius and Hertzsprung-Russell diagrams compared to isochrones for ages of  6.8\,Gyr assuming $[{\rm Fe/H}]=-0.2$ from the Dartmouth stellar evolution database (cyan dotted line) and MIST (green dashed line). The ellipses show 1-$\sigma$ and  2-$\sigma$ confidence regions on the parameters of EBLM\,J0113+31\,B. Parameters for very low mass stars shown as error bars in blue are taken from DEBCat \citep{2015ASPC..496..164S}.}
    \label{fig:isochrones2}
\end{figure}

The same isochrones from the Dartmouth and MIST stellar model grids are compared to the properties of EBLM~J0113+31\,B in Fig.~\ref{fig:isochrones2}.  Our grid of {\sc garstec} models does not extend to these very low masses. The agreement between the models and observations is reasonably good, which is somewhat surprising given the long-standing observation that  stellar models tend to under-predict the radius and over-predict $T_{\rm eff}$ for low-mass stars \citep{2013ApJ...776...87S, 2019A&A...626A..32C, 2014MNRAS.437.2831Z, 2006ApJ...644..475B, 1973A&A....26..437H, 1977ApJS...34..479L}. This can be seen from the mass, radius and  $T_{\rm eff}$ measurements for six other very low-mass stars in the same figure. These six stars are members of three eclipsing binaries with orbital periods less 2 days. This complicates the interpretation of their properties in the light of the so-called ``radius inflation'' problem because these stars will be forced to rotate much faster than most single M-dwarf stars by tidal forces in these short-period binaries. EBLM~J0113+31\,B is a valuable addition to the small sample of well-characterised VLMSs because we have an independent estimate of its age and initial metal abundance based on observations of the G0V primary star to add to the accurate mass, radius and $T_{\rm eff}$ measurements.

\subsection{EBLM systems as benchmark stars}

Benchmark FGK dwarf stars with direct $T_{\rm eff}$ measurements based on angular diameters measured by interferometry typically have apparent magnitudes ${\rm V}=1$\,--\,6 \citep{2014A&A...564A.133J}. This is 5\,--\,10 magnitudes brighter than the magnitude limits for large-scale spectroscopic surveys, so special observing modes must be employed to obtain spectra of these benchmark stars. These bright benchmark stars also tend to be single stars, so there are often no direct measurements of 
their mass or surface gravity. It is difficult to extend this sample because new candidates for benchmark stars will necessarily be more distant than the existing sample, i.e. they will have smaller angular diameters than the existing benchmark stars. For example, a nominal Sun-like star at distance of 10\,pc will have an angular diameter $\theta = 0.465$\,mas, so a systematic error of only 0.04\,mas, which is typical for existing measurements \citep{2022A&A...658A..47K}, implies a systematic error of 250\,K in the measured value of $T_{\rm eff}$ for such a star.

In contrast, EBLM~J0113+31 is within the magnitude range of recent large-scale spectroscopic surveys, e.g. the TESS-HERMES survey \citep 
[$10 < {\rm V} < 13.1$,][]{2018MNRAS.473.2004S}, LAMOST ``VB mode'' observations \citep 
[$9.0 \le J \le 12.5$,][]{2015RAA....15.1095L}, and stars in open clusters observed as part of the Gaia-ESO survey \citep
[$9 < {\rm V}< 16.5$,][]{2021arXiv211211974B}. 
This makes it feasible to observe EBLM\,J0113+31 and other EBLM binaries in exactly the same way as other stars observed by these survey instruments as part of their routine operations. The contribution of the M-dwarf to the total flux at optical wavelengths is $\loa 0.2$\,per~cent for EBLM binaries, so the M-dwarf will have a completely negligible effect on the atmospheric parameters derived from the analysis of the optical spectrum. This makes it possible to make an ``end-to-end'' test of the accuracy of parameters derived by the combination of these survey instruments plus their data processing and analysis pipelines. Even at near-infrared wavelengths used by surveys such as APOGEE \citep{2018AJ....156..126J} the contribution from the M-dwarf is $\loa  1$\,per~cent, so the results of any analysis that includes a correction for this small contribution to the total flux will be insensitive to the details of how this correction is done. 

 Many EBLM binaries in the magnitude range $10\loa{\rm V}\loa12$ have been identified and have well-determined spectroscopic orbits that have been published \citep{2017A&A...608A.129T} or that are in preparation thanks to the EBLM project and BEBOP survey \citep{2022MNRAS.tmp..155S}. High-quality space-based photometry is already available for many of these stars from the TESS survey and/or from our on-going CHEOPS GTO programme. Several échelle spectrographs that can provide high-resolution spectroscopy at near-infrared wavelengths are currently operational on 4\,--\,10\,m telescopes, e.g. CARMENES on the Calar Alto Observatory 3.5-m telescope \citep{2016SPIE.9908E..12Q}, NIRPS on the ESO 3.6-m telescope \citep{2021EPSC...15..555G}, CRIRES+ on the ESO 8.2-m VLT \citep{2004SPIE.5492.1218K}, and IRD on the 8.2-m Subaru telescope \citep{2014SPIE.9147E..14K}. We can also look forward to  high-quality spectrophotometry and improved parallax measurements for these EBLM systems in future data releases from the Gaia mission.\footnote{\url{https://www.cosmos.esa.int/web/gaia/release}} In summary, the instrumentation, data and targets needed to create a network of moderately-bright FGK dwarf stars covering both hemispheres that are ideal benchmark stars for on-going large-scale spectroscopic surveys are all now available.
 
 Apart from their utility as benchmarks stars for large-scale spectroscopic surveys, follow-up observations of additional EBLM systems will also provide valuable data on the properties of very low-mass stars. With observations similar to those presented here we can create a sample VLMSs with precise and accurate $T_{\rm eff}$, mass and radius measurements. These EBLM binaries will have independent estimates for their age and initial metallicity based on the observed properties of the primary stars in these systems. It is not feasible to obtain a direct spectrum for these very faint companion stars, but it should be possible given sufficiently high-quality data to estimate the projected rotational velocity of the star from the width of the peak in the stacked-CCF. Data of this quality will be very useful for testing and calibrating models of very low mass stars that include additional physics to account for the radius inflation problem \citep{2018ApJ...869..149M, 2014A&A...571A..70F}.
 
 Many of these EBLM binary systems will also be ideal benchmark stars for the upcoming PLATO mission \citep{2014ExA....38..249R} if we can measure model-independent masses for the primary star using the techniques presented in this study. The PLATO mission will focus on bright stars  (4\,--\,11 mag)  with the aim to detect and characterize planets down to Earth-size by photometric transits. Asteroseismology will be performed for these bright stars to obtain stellar parameters, including masses and ages. The PLATO Definition Study Report\footnote{\url{https://sci.esa.int/science-e/www/object/doc.cfm?fobjectid=59251}} (``red book'') specifies that PLATO must be capable of delivering accurate stellar ages with a precision of 10\,per~cent. Some corrections for systematic errors in the current generation of stellar models will be needed to reach this accuracy in stellar ages \citep{2017EPJWC.16001003G}. The planned observing strategy includes a step-and-stare phase that will cover about 50\,per-cent of the sky. EBLM binaries can be used to perform  ``end-to-end'' tests of the PLATO data analysis to ensure that the mass estimates delivered for these stars are accurate, and to calibrate the next generation of stellar models using direct mass, radius, and $T_{\rm eff}$ measurements combined with asteroseismology.

\section{Conclusions}

We have derived precise and accurate masses, radii and effective temperatures for both stars in the eclipsing binary system EBLM~J0113+31.  These data can be used to validate and calibrate stellar models, empirical relations for stellar properties, and to test data analysis techniques. With the techniques established here, it is feasible to create a network of moderately-bright FGK dwarf stars covering both hemispheres that are ideal benchmarks for on-going large-scale spectroscopic surveys and for the upcoming PLATO mission.

\section*{Acknowledgements}

CHEOPS is an ESA mission in partnership with Switzerland with important contributions to
the payload and the ground segment from Austria, Belgium, France, Germany, Hungary, Italy,
Portugal, Spain, Sweden, and the United Kingdom. The CHEOPS Consortium would like to
gratefully acknowledge the support received by all the agencies, offices, universities, and
industries involved. Their flexibility and willingness to explore new approaches were essential
to the success of this mission.

Based on observations obtained at the Canada-France-Hawaii Telescope (CFHT) which is operated from the summit of Maunakea by the National Research Council of Canada, the Institut National des Sciences de l'Univers of the Centre National de la Recherche Scientifique of France, and the University of Hawaii. The observations at the Canada-France-Hawaii Telescope were performed with care and respect from the summit of Maunakea which is a significant cultural and historic site. Based on observations obtained with SPIRou, an international project led by Institut de Recherche en Astrophysique et Plan\'{e}tologie, Toulouse, France. PM is grateful to the observatory staff at the CHFT for their help with the planning, execution and data reduction of the SPIRou data used for this study. 

Based on observations made with the Nordic Optical Telescope, owned in collaboration by the University of Turku and Aarhus University, and operated jointly by Aarhus University, the University of Turku and the University of Oslo, representing Denmark, Finland and Norway, the University of Iceland and Stockholm University at the Observatorio del Roque de los Muchachos, La Palma, Spain, of the Instituto de Astrofisica de Canarias.

This paper includes data collected by the TESS mission, which is publicly available from the Mikulski Archive for Space Telescopes (MAST) at the Space Telescope Science Institure (STScI). Funding for the TESS mission is provided by the NASA Explorer Program directorate. STScI is operated by the Association of Universities for Research in Astronomy, Inc., under NASA contract NAS 5–26555. We acknowledge the use of public TESS Alert data from pipelines at the TESS Science Office and at the TESS Science Processing Operations Center.

The authors would like to thank the anonymous referee for their useful comments that have helped to improve this manuscript.

This research made use of Lightkurve, a Python package for Kepler and TESS data analysis \citep{2018ascl.soft12013L}.

PM acknowledges support from UK Science and Technology Facilities Council (STFC) research grant number ST/M001040/1.

NM is supported by STFC grant number ST/S505444/1.

SH gratefully acknowledges CNES funding through the grant 837319.

S.G.S. acknowledge support from FCT through FCT contract nr. CEECIND/00826/2018 and POPH/FSE (EC).

ACC acknowledges support from STFC consolidated grant numbers ST/R000824/1 and ST/V000861/1, and UKSA grant number ST/R003203/1.

YA and MJH acknowledge the support of the Swiss National Fund under grant 200020\_172746.

MIS acknowledges support from STFC grant number ST/T506175/1.

We acknowledge support from the Spanish Ministry of Science and Innovation and the European Regional Development Fund through grants ESP2016-80435-C2-1-R, ESP2016-80435-C2-2-R, PGC2018-098153-B-C33, PGC2018-098153-B-C31, ESP2017-87676-C5-1-R, MDM-2017-0737 Unidad de Excelencia Maria de Maeztu-Centro de Astrobiología (INTA-CSIC), as well as the support of the Generalitat de Catalunya/CERCA programme. The MOC activities have been supported by the ESA contract No. 4000124370.

S.C.C.B. acknowledges support from FCT through FCT contracts nr.IF/01312/2014/CP1215/CT0004.

XB, SC, DG, MF and JL acknowledge their role as ESA-appointed CHEOPS science team members.

ABr was supported by the SNSA.

This project was supported by the CNES.

The Belgian participation to CHEOPS has been supported by the Belgian Federal Science Policy Office (BELSPO) in the framework of the PRODEX Program, and by the University of Li\`{e}ge through an ARC grant for Concerted Research Actions financed by the Wallonia-Brussels Federation.

L.D. is an F.R.S.-FNRS Postdoctoral Researcher.

This work was supported by FCT - Funda\c{c}\~{a}o para a Ciência e a Tecnologia through national funds and by FEDER through COMPETE2020 - Programa Operacional Competitividade e Internacionalizac\~{a}o by these grants: UID/FIS/04434/2019, UIDB/04434/2020, UIDP/04434/2020, PTDC/FIS-AST/32113/2017 \& POCI-01-0145-FEDER- 032113, PTDC/FIS-AST/28953/2017 \& POCI-01-0145-FEDER-028953, PTDC/FIS-AST/28987/2017 \& POCI-01-0145-FEDER-028987, O.D.S.D. is supported in the form of work contract (DL 57/2016/CP1364/CT0004) funded by national funds through FCT.

B.-O.D. acknowledges support from the Swiss National Science Foundation (PP00P2-190080).
This project has received funding from the European Research Council (ERC) under the European Union’s Horizon 2020 research and innovation programme (project {\sc Four Aces}.
grant agreement No 724427). It has also been carried out in the frame of the National Centre for Competence in Research PlanetS supported by the Swiss National Science Foundation (SNSF). DE acknowledges financial support from the Swiss National Science Foundation for project 200021\_200726.

MF and CMP gratefully acknowledge the support of the Swedish National Space Agency (DNR 65/19, 174/18).

DG gratefully acknowledges financial support from the CRT foundation under Grant No. 2018.2323 ``Gaseous or rocky? Unveiling the nature of small worlds''.

M.G. is an F.R.S.-FNRS Senior Research Associate.

KGI is the ESA CHEOPS Project Scientist and is responsible for the ESA CHEOPS Guest Observers Programme. She does not participate in, or contribute to, the definition of the Guaranteed Time Programme of the CHEOPS mission through which observations described in this paper have been taken, nor to any aspect of target selection for the programme.

This work was granted access to the HPC resources of MesoPSL financed by the Region Ile de France and the project Equip@Meso (reference ANR-10-EQPX-29-01) of the programme Investissements d'Avenir supervised by the Agence Nationale pour la Recherche.

ML acknowledges support of the Swiss National Science Foundation under grant number PCEFP2\_194576.

GSc, GPi, IPa, LBo, VNa and RRa acknowledge the funding support from Italian Space Agency (ASI) regulated by “Accordo ASI-INAF n. 2013-016-R.0 del 9 luglio 2013 e integrazione del 9 luglio 2015 CHEOPS Fasi A/B/C”.

This work was also partially supported by a grant from the Simons Foundation (PI Queloz, grant number 327127).

IR acknowledges support from the Spanish Ministry of Science and Innovation and the European Regional Development Fund through grant PGC2018-098153-B- C33, as well as the support of the Generalitat de Catalunya/CERCA programme.

GyMSz acknowledges the support of the Hungarian National Research, Development and Innovation Office (NKFIH) grant K-125015, a PRODEX Institute Agreement between the ELTE E\"otv\"os Lor\'and University and the European Space Agency (ESA-D/SCI-LE-2021-0025), the Lend\"ulet LP2018-7/2021 grant of the Hungarian Academy of Science and the support of the city of Szombathely.

V.V.G. is an F.R.S-FNRS Research Associate.

%%%%%%%%%%%%%%%%%%%%%%%%%%%%%%%%%%%%%%%%%%%%%%%%%%
\section*{Data Availability}
The data underlying this article are available in the following repositories: 
 The Data \& Analysis Center for Exoplanets (DACE) -- \url{https://dace.unige.ch/} (CHEOPS);
Canadian Astronomy Data Centre -- \url{https://www.cadc-ccda.hia-iha.nrc-cnrc.gc.ca/} (CFHT/SPIRou); 
Mikulski Archive for Space Telescopes -- \url{https://archive.stsci.edu/}  (TESS).



\bibliographystyle{mnras}
\bibliography{allbib} % if your bibtex file is called example.bib

%%%%%%%%%%%%%%%%%%%%%%%%%%%%%%%%%%%%%%%%%%%%%%%%%%

\bigskip
% List of institutions
\noindent
$^{1}$ Astrophysics Group, Keele University, Keele, Staffordshire ST5 5BG, UK \\
$^{2}$ Aix Marseille Univ, CNRS, CNES, LAM, 38 rue Frédéric Joliot-Curie, 13388 Marseille, France \\
$^{3}$ Instituto de Astrofisica e Ciencias do Espaco, Universidade do Porto, CAUP, Rua das Estrelas, 4150-762 Porto, Portugal \\
$^{4}$ Observatoire Astronomique de l'Université de Genève, Chemin Pegasi 51, Versoix, Switzerland \\
$^{5}$ Physikalisches Institut, University of Bern, Gesellsschaftstrasse 6, 3012 Bern, Switzerland \\
$^{6}$ Center for Space and Habitability, Gesellsschaftstrasse 6, 3012 Bern, Switzerland \\
$^{7}$ Centre for Exoplanet Science, SUPA School of Physics and Astronomy, University of St Andrews, North Haugh, St Andrews KY16 9SS, UK \\
$^{8}$ Division Technique INSU, CS20330, 83507 La Seyne sur Mer cedex, France \\
$^{9}$ School of Physics and Astronomy, University of Birmingham, Edgbaston, Birmingham B15 2TT, UK \\
$^{10}$ Instituto de Astrofisica de Canarias, 38200 La Laguna, Tenerife, Spain \\
$^{11}$ Departamento de Astrofisica, Universidad de La Laguna, 38206 La Laguna, Tenerife, Spain \\
$^{12}$ Institut de Ciencies de l'Espai (ICE, CSIC), Campus UAB, Can Magrans s/n, 08193 Bellaterra, Spain \\
$^{13}$ Institut d'Estudis Espacials de Catalunya (IEEC), 08034 Barcelona, Spain \\
$^{14}$ Admatis, 5. Kandó Kálmán Street, 3534 Miskolc, Hungary \\
$^{15}$ Depto. de Astrofisica, Centro de Astrobiologia (CSIC-INTA), ESAC campus, 28692 Villanueva de la Cañada (Madrid), Spain \\
$^{16}$ Departamento de Fisica e Astronomia, Faculdade de Ciencias, Universidade do Porto, Rua do Campo Alegre, 4169-007 Porto, Portugal \\
$^{17}$ Space Research Institute, Austrian Academy of Sciences, Schmiedlstrasse 6, A-8042 Graz, Austria \\
$^{18}$ Université Grenoble Alpes, CNRS, IPAG, 38000 Grenoble, France \\
$^{19}$ Department of Astronomy, Stockholm University, AlbaNova University Center, 10691 Stockholm, Sweden \\
$^{20}$ Institute of Planetary Research, German Aerospace Center (DLR), Rutherfordstrasse 2, 12489 Berlin, Germany \\
$^{21}$ Université de Paris, Institut de physique du globe de Paris, CNRS, F-75005 Paris, France \\
$^{22}$ ESTEC, European Space Agency, 2201AZ, Noordwijk, NL \\
$^{23}$ Centre for Mathematical Sciences, Lund University, Box 118, 221 00 Lund, Sweden \\
$^{24}$ Astrobiology Research Unit, Université de Liège, Allée du 6 Août 19C, B-4000 Liège, Belgium \\
$^{25}$ Space sciences, Technologies and Astrophysics Research (STAR) Institute, Université de Liège, Allée du 6 Août 19C, 4000 Liège, Belgium \\
$^{26}$ Leiden Observatory, University of Leiden, PO Box 9513, 2300 RA Leiden, The Netherlands \\
$^{27}$ Department of Space, Earth and Environment, Chalmers University of Technology, Onsala Space Observatory, 43992 Onsala, Sweden \\
$^{28}$ Dipartimento di Fisica, Universita degli Studi di Torino, via Pietro Giuria 1, I-10125, Torino, Italy \\
$^{29}$ University of Vienna, Department of Astrophysics, Türkenschanzstrasse 17, 1180 Vienna, Austria \\
$^{30}$ Department of Physics, University of Warwick, Gibbet Hill Road, Coventry CV4 7AL, United Kingdom \\
$^{31}$ Airbus defence and Space, Avenida de Aragon 404, 28022 Madrid, Spain \\
$^{32}$ Science and Operations Department - Science Division (SCI-SC), Directorate of Science, European Space Agency (ESA), European Space Research and Technology Centre (ESTEC),
Keplerlaan 1, 2201-AZ Noordwijk, The Netherlands \\
$^{33}$ Konkoly Observatory, Research Centre for Astronomy and Earth Sciences, 1121 Budapest, Konkoly Thege Miklós út 15-17, Hungary \\
$^{34}$ ELTE E\"{o}tv\"{o}s Lor\'{a}nd University, Institute of Physics, P\'{a}zm\'{a}ny P\'{e}ter s\'{e}t\'{a}ny 1/A, 1117 Budapest, Hungary \\
$^{35}$ IMCCE, UMR8028 CNRS, Observatoire de Paris, PSL Univ., Sorbonne Univ., 77 av. Denfert-Rochereau, 75014 Paris, France \\
$^{36}$ Institut d'astrophysique de Paris, UMR7095 CNRS, Université Pierre \& Marie Curie, 98bis blvd. Arago, 75014 Paris, France \\
$^{37}$ INAF, Osservatorio Astronomico di Padova, Vicolo dell'Osservatorio 5, 35122 Padova, Italy \\
$^{38}$ INAF, Osservatorio Astrofisico di Catania, Via S. Sofia 78, 95123 Catania, Italy \\
$^{39}$ Department of Astrophysics, University of Vienna, Tuerkenschanzstrasse 17, 1180 Vienna, Austria \\
$^{40}$ Institute of Optical Sensor Systems, German Aerospace Center (DLR), Rutherfordstrasse 2, 12489 Berlin, Germany \\
$^{41}$ Dipartimento di Fisica e Astronomia "Galileo Galilei", Universita degli Studi di Padova, Vicolo dell'Osservatorio 3, 35122 Padova, Italy \\
$^{42}$ ETH Zurich, Department of Physics, Wolfgang-Pauli-Strasse 2, CH-8093 Zurich, Switzerland \\
$^{43}$ Cavendish Laboratory, JJ Thomson Avenue, Cambridge CB3 0HE, UK \\
$^{44}$ Center for Astronomy and Astrophysics, Technical University Berlin, Hardenberstrasse 36, 10623 Berlin, Germany \\
$^{45}$ Institut für Geologische Wissenschaften, Freie Universität Berlin, 12249 Berlin, Germany \\
$^{46}$ MTA-ELTE Exoplanet Research Group, 9700 Szombathely, Szent Imre h. u. 112, Hungary \\
$^{47}$ Institute of Astronomy, University of Cambridge, Madingley Road, Cambridge, CB3 0HA, United Kingdom \\ 

%\begin{thebibliography}{99}
%\end{thebibliography}

%%%%%%%%%%%%%%%%%%%%%%%%%%%%%%%%%%%%%%%%%%%%%%%%%%

%%%%%%%%%%%%%%%%% APPENDICES %%%%%%%%%%%%%%%%%%%%%

%\appendix

%\section{Radial velocity measurements}
%Radial velocity measurements for EBLM~J0113+31\,A measured from the SPIRou spectra of %EBLM~J0113+31 are given in Table~\ref{table:rv}.

% Don't change these lines
\bsp	% typesetting comment
\label{lastpage}
\end{document}